\documentclass[useAMS,usenatbib]{mn2e}

\usepackage{epsfig}
\usepackage{graphicx}
\usepackage{journal_names}
\usepackage[english]{babel}

\usepackage{amsmath}
\usepackage{amssymb}
\usepackage{amsthm}
\usepackage{fancyhdr,txfonts} 

\usepackage{natbib}
\usepackage{sidecap}
\usepackage{subfig}

\usepackage{color}
\usepackage{wasysym}
\usepackage{url}

\makeatletter
\newcommand\footnoteref[1]{\protected@xdef\@thefnmark{\ref{#1}}\@footnotemark}
\makeatother

\title[Morphology of Red Sequence Galaxies]{The Morphological Transformation of Red Sequence Galaxies in Clusters since $z \sim 1$}
\author[P. Cerulo et al.]{\parbox{\textwidth}{P. Cerulo$^{1,2}$\thanks{E-mail:pcerulo@astro-udec.cl}, W. J. Couch{$^{2,3}$}, C. Lidman$^{3}$, R. Demarco{$^{1}$}, M. Huertas-Company{$^{4}$}, S. Mei{$^{4}$}, R. S\'{a}nchez-Janssen{$^{5}$}, L. F. Barrientos{$^{6,7}$}, R. Mu\~{n}oz{$^{7}$}}\vspace{0.4cm}\\
\parbox{\textwidth}{$^{1}$Department of Astronomy, Universidad de Concepci\'{o}n, Casilla 160-C Concepci\'{o}n, Chile\\
$^{2}$Centre for Astrophysics and Supercomputing, Swinburne University of Technology, PO Box 218, Hawthorn, VIC 3122, Australia\\
$^{3}$Australian Astronomical Observatory, PO Box 915, North Ryde, NSW 1670, Australia \\
$^{4}$GEPI, Paris Observatory, 77 av. Denfert Rochereau, 75014 Paris, France\\
$^{5}$UK Astronomy Technology Centre, Royal Observatory Edinburgh, Blackford Hill, Edinburgh, EH9 3HJ, UK\\
$^{6}$Instituto de Astrofisica, Pontificia Universidad Cat\'{o}lica de Chile\\
$^{7}$Millennium Institute of Astrophysics, Chile\\}}
\begin{document}


\pagerange{\pageref{firstpage}--\pageref{lastpage}} \pubyear{2002}

\maketitle

\label{firstpage}

\begin{abstract}
The study of galaxy morphology is fundamental to understand the physical processes driving the structural evolution of galaxies. It has long been known that dense environments host high fractions of early-type galaxies and low fractions of late-type galaxies, indicating that the environment affects the structural evolution of galaxies. In this paper we present an analysis of the morphological composition of red sequence galaxies in a sample of 9 galaxy clusters at $0.8<z<1.5$ drawn from the HAWK-I cluster survey (HCS), with the aim of investigating the evolutionary paths of galaxies with different morphologies. We classify galaxies according to their apparent bulge-to-total light ratio and compare with red sequence galaxies from the lower-redshift WINGS and EDisCS surveys. We find that, while the HCS red sequence is dominated by elliptical galaxies at all luminosities and stellar masses, the WINGS red sequence is dominated by elliptical galaxies only at its bright end ($M_V<-21.0$ mag), while S0s become the most frequent class at fainter luminosities. Disc-dominated galaxies comprise 10-14\% of the red sequence population in the low (WINGS) and high (HCS) redshift samples, although their fraction increases up to 40\% at $0.4 < z < 0.8$ (EDisCS). We find a 20\% increase in the fraction of S0 galaxies from $z \sim 1.5$ to $z \sim 0.05$ on the red sequence. These results suggest that elliptical and S0 galaxies follow different evolutionary histories and, in particular, that S0 galaxies result, at least at intermediate luminosities ($-22.0 < M_V < -20.0$), from the morphological transformation of quiescent spiral galaxies.
\end{abstract}


\section{Introduction}

Clusters of galaxies are the most massive virialised large scale structures in the universe. They host a variety of environments, from the dense and virialised cores to the sparse outskirts and infall regions, making them suitable for studies of the environmental drivers of galaxy evolution.

A morphology vs density relation, whereby the fraction of elliptical and S0 (early-type) galaxies increases towards regions with higher galaxy density, while the fraction of spiral and irregular (late-type) galaxies becomes higher at low galaxy densities, is observed in rich clusters up to redshift $z=1.5$ (\citealt{Dressler_1980}, \citealt{Dressler_1997}, \citealt{Postman_2005}, \citealt{Holden_2007}, \citealt{van_der_wel_2007}, \citealt{Mei_2012}). \textcolor{black}{In the same redshift range is also observed a star-formation vs density relation whereby old and passively evolving galaxies dominate the highest density regions while young and star-forming systems become more frequent in sparse regions (see e.g.: \citealt{Ellingson_2001}, \citealt{Fritz_2014}).} The simultaneous existence of these two correlations implies that the cluster environment must play an important role in determining the evolutionary pathways of galaxies, setting the timescales for the growth of the red sequence and influencing structural evolution. Thus the study of the evolution of the red sequence in clusters should not be separated from the study of the evolution of galaxy morphologies as the two aspects are strongly connected. In particular, it is crucial to understand the timescales for the transformations in the morphology and stellar populations of galaxies as these may be significantly different, and one transformation may influence the other as shown in many theoretical models.

Since the work of E. P. Hubble in 1926 \citep{Hubble_1926}, galaxies have been qualitatively categorised in two main classes of elliptical and spiral, with a transition lenticular or S0 class characterised by the presence of a bulge and a disc, as observed in spiral galaxies, but without spiral arms, as observed in elliptical galaxies. Galaxies not showing any elliptical or spiral feature but only irregular/undefined morphologies (e.g.\ the Magellanic Clouds) were classified as irregular. 

The evolutionary links between galaxies with different morphologies are one of the main open questions in modern astrophysics. The observations have shown that elliptical and S0 galaxies have older stellar populations than spiral and irregular galaxies. Furthermore, elliptical and S0 galaxies host little or no ongoing star formation, making them the main constituent of the red sequence. These similarities have motivated astronomers to study together, on one hand, ellipticals and S0s as one class of {\itshape{early-type}} galaxies and, on the other hand, spirals and irregulars as one class of {\itshape{late-type}} galaxies. The differences between stellar populations suggest that galaxies were born spirals and then evolved through subsequent phases into S0s and ellipticals (see e.g.\ \citealt{van_Dokkum_1998}, \citealt{Tran_2007}, \citealt{Sanchez_Blazquez_2009}), although the very old stellar populations found in some elliptical galaxies suggest very short formation timescales not necessarily implying a transformation from spirals (see e.g.\ \citealt{Thomas_2005}).

As discussed in \cite{Toomre_1977} and \cite{Ghosh_2015}, spiral arms in galaxy discs do not survive for more than $\sim$1 Gyr without the support of the interstellar gas, which slows down the velocity of the density wave packets giving birth to spiral arms. As a result, all physical mechanisms causing gas depletion in spiral galaxies may promote transformations into S0 or elliptical morphologies. For instance, bulge growth triggering starbursts towards the galaxy centre and heating the gas in the disc can be responsible for such a transition in low-density environments (e.g.\ \citealt{Somerville_2008}, \citealt{Martig_2009}). On the other hand, processes such as mergers \citep{Lavery_1988}, harassment \citep{Moore_1998}, tidal interactions as galaxies fly by each other \citep{Bekki_2011}, strangulation \citep{Larson_1980}, and ram-pressure stripping  (\citealt{Gunn_1972}, \citealt{Bekki_2014}) may all conduct to such transformations in dense environments (see \citealt{Treu_2003} for a review). Although these processes are all effective in theoretical models, it is not yet clear from the observations what is the main driver of morphological evolution and what role the environment plays in determining the pathways of morphological transformations.

Although the morphology - star-formation - density relation has been observed to be in place since at least $z=1.5$, \cite{Dressler_1997} and \cite{Postman_2005} remarked that, while the fraction of elliptical galaxies follows the same trend with local density in clusters at all redshifts, the fraction of S0 galaxies decreases \textcolor{black}{towards higher redshifts}, with a corresponding increase in the fraction of late-type galaxies. This conclusion is in agreement with results showing a mild variation in the fraction of elliptical galaxies over redshift and an increase in the fraction of S0 galaxies towards low redshifts in both clusters (\citealt{Couch_1998}, \citealt{Fasano_2001}, \citealt{Holden_2007}, \citealt{van_der_wel_2007}, \citealt{Vulcani_2011}, \citealt{Donofrio_2015}) and the field (\citealt{Bundy_2005}, \citealt{Oesch_2010}, \citealt{Huertas_2015a, Huertas_2016}) up to $z \sim 3$. All these works suggest that, despite the similarities in stellar age, elliptical and S0 galaxies follow two distinct evolutionary paths and, in particular, that spiral galaxies may be the progenitors of S0 galaxies.

\cite{Poggianti_2006} proposed an evolutionary scenario in which cluster early-type galaxies are made up of two populations, a pristine population coeval to the cluster, and a quenched population probably resulting mostly from the morphological transformation of quiescent spiral galaxies on the red sequence ({\itshape{red spirals}}, \citealt{Wolf_2009}, \citealt{Masters_2010}). Such a scenario is in agreement with studies of the stellar populations in galaxies with different morphologies in clusters. For instance, as shown in \cite{Poggianti_2001}, the average stellar age of low-mass S0 galaxies in the Coma cluster is lower than both elliptical and higher mass S0s (see also \citealt{van_Dokkum_1998}). \cite{Tran_2007} showed that red sequence S0s in the cluster MS 1054-03, at $z=0.83$, are younger than ellipticals but older than late-type galaxies. \cite{Mei_2009} showed that S0 galaxies on the red sequence of the 8 clusters of the Advanced Camera for Surveys (ACS) Intermediate Redshift Cluster Survey \citep{Ford_2004}, at $0.8< z < 1.3$, are younger than ellipticals at the same luminosities, while at lower luminosities the ages of elliptical and S0 galaxies are similar. These results all support the notion of different evolutionary histories for elliptical and S0s, with the latter being on average younger.

In \cite{Cerulo_2016} we presented an analysis of the properties of the red sequence in a sample of clusters at $0.8<z<1.5$ drawn from the HAWK-I Cluster Survey (HCS, \citealt{Lidman_2013}), finding that halo mass plays a major role in setting the timescales for the growth of the red sequence. In particular, we found that the red sequence in clusters is more evolved at its faint end with respect to the field, suggesting that the effect of the environment is to accelerate star-formation quenching in galaxies, in agreement with other works in the literature at similar redshifts (see e.g.: \citealt{Gobat_2008}, \citealt{Rettura_2011}, \citealt{Muzzin_2012}). In this paper, which is the second in the series of red sequence papers in the HCS started with \cite{Cerulo_2016}, we present an analysis of the morphological composition of the red sequence in the HCS with the aim of studying the evolutionary histories of galaxies with different morphological types and establishing the links between the build-up of the red sequence and the structural evolution of galaxies.

The paper is organised as follows. In Section 2 we describe the observations and measurements; Section 3 presents the results, which are then discussed in Section 4. Section 5 finally summarises the results and draws the main conclusions of this investigation.

Throughout the paper we adopt a $\Lambda CDM$ cosmology with $\Omega_\Lambda = 0.73$, $\Omega_{m} = 0.27$, and $H_0 = 71.0$ $km \cdot s^{-1} \cdot Mpc^{-1}$. Unless otherwise stated, all magnitudes are quoted in the AB system \citep{Oke_1974}. This is the same convention followed in \cite{Cerulo_2014} and \cite{Cerulo_2016}. We define $R_{200}$ as the physical radius, measured in Mpc, including the region where the total matter density (baryonic and non-baryonic) is 200 times higher than the critical density at the redshift of each cluster.

\begin{figure*}
	\subfloat[]{\label{fig:a}}\includegraphics[width=0.8\textwidth, trim=0.0cm 2.5cm 0.0cm 0.0cm, clip, page=1]{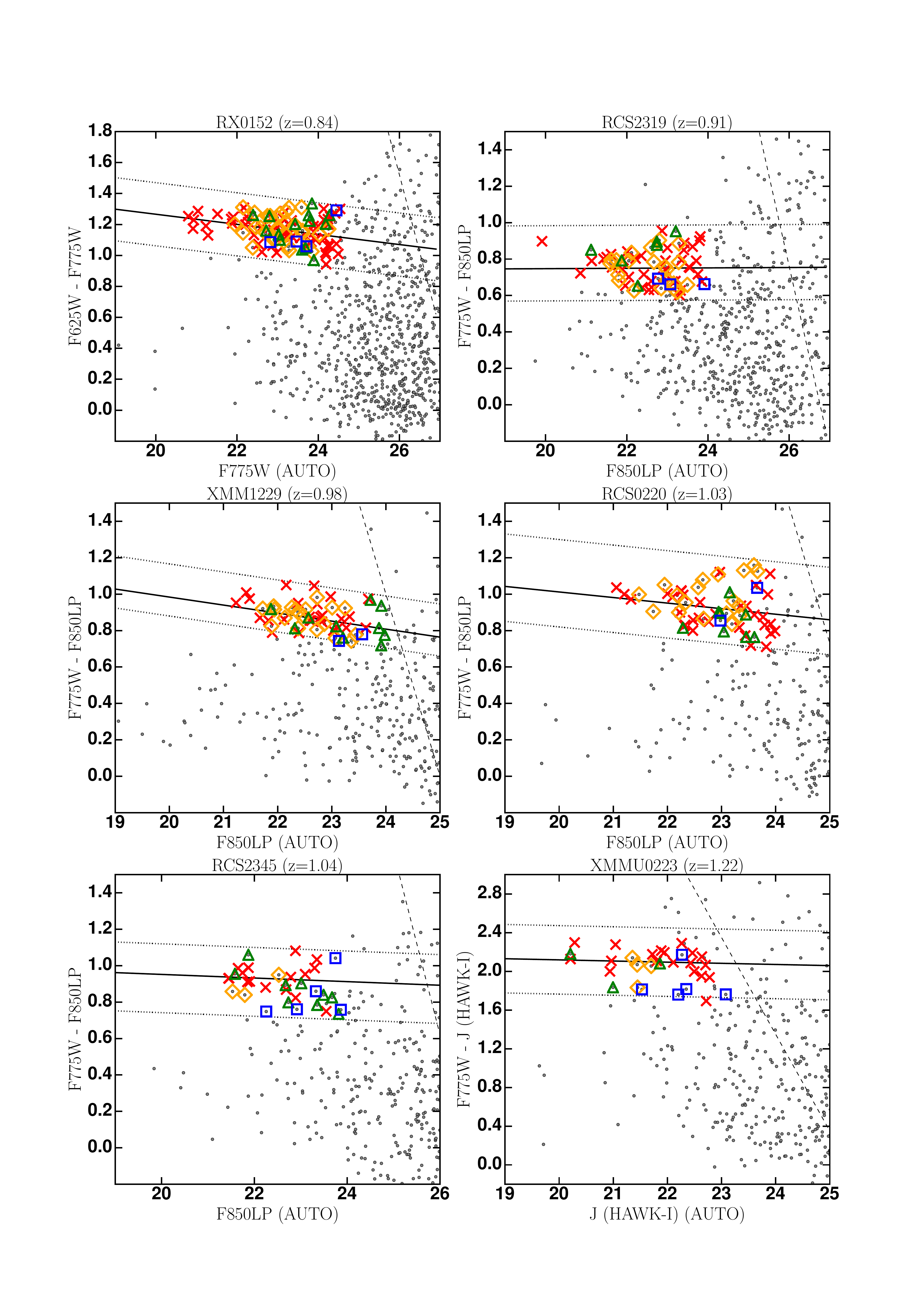}
	\caption{Observed colour-magnitude diagrams of the individual HCS clusters with the morphologically classified red sequence galaxies highlighted by different colours and symbols. All the red sequence galaxies with $z_{850} < 24.0$ mag were selected for morphological classification. Grey points are all galaxies observed in the field of each cluster within $0.54 \times R_{200}$ from the cluster centroid. Red crosses are elliptical galaxies (E), orange diamonds are bulge-dominated/S0 galaxies (BD), green triangles are early-type disc-dominated galaxies (EDD), and blue squares are late-type disc-dominated and irregular galaxies (LDD+Irr).  The solid lines represent the fit to the observed red sequence. The dotted lines mark the boundaries of the red sequence. The diagonal dashed lines represent the 90\% magnitude completeness limits. As it can be seen, due to the magnitude cut $z_{850} = 24.0$, the red sequence is not covered down to the faint end in all the clusters and in some cases, as in XMMXCS2215 (bottom-left panel on next page) only galaxies which are bluer than the best-fit line fall in the sample.}
\end{figure*}

\begin{figure*}
  \ContinuedFloat 
  \centering 
  \subfloat[]{\label{fig:b}}\includegraphics[width=0.8\textwidth, trim=0.0cm 18.5cm 0.0cm 0.0cm, clip, page=2]{plot_HCS_CM_morphology}
  \caption{{\itshape{(Continued)}}.}
\label{fig1}
\end{figure*}

\section{Observations and Measurements}

\subsection{The HAWK-I Cluster Survey}

The sample used in this paper is composed of 9 clusters of galaxies at $0.8 < z < 1.5$ drawn from the HAWK-I Cluster Survey (HCS, \citealt{Lidman_2013}) and observed in at least 4 pass bands from $R$ to $Ks$. The sample is presented in \cite{Cerulo_2016}, and we refer to that paper for a detailed discussion of the observations and data reduction and for a description of the method adopted in the photometry. In this paper we report a summary of the cluster sample in Table \ref{table1}.

All the clusters were observed in the Hubble Space Telescope Advanced Camera for Surveys (HST/ACS) F850LP ($z_{850}$) band, which delivered deep ($z_{850} = 25-27$ mag 90\% completeness limit) and high resolution images with pixel scale $0.05''$/pixel and image quality, parametrised by the Full Width at Half Maximum (FWHM) of the Point Spread Function (PSF), $FWHM \sim 0.09''$ in all images. We used these images to classify red sequence galaxies, while we used the data taken in all the other filters shown in Table 2 of \cite{Cerulo_2016} to estimate stellar masses.

The numbers of spectroscopically confirmed members in the HCS clusters vary from cluster to cluster in the range 18-130. For this work we used all the galaxies observed within a radius $0.54 \times R_{200}$ from each cluster centre. This allowed us to increase the number of galaxies in our sample, although we had to implement a statistical technique to take into account the contamination of background galaxies as explained in Section 3.1. 

For each HCS cluster we considered the region within $0.54 \times R_{200}$ from the cluster centre and modelled the red sequence in the observer frame with a linear relationship parametrised in terms of its zero-point, slope and intrinsic scatter. The red sequence population was defined as those galaxies in the colour range:
\begin{equation}\label{eq1}
-\kappa_l \sigma_{c} < \left( C - C_{RS} \right) < +\kappa_h \sigma_{c}
\end{equation}
where $C$ is the observed colour, $C_{RS}$ is the colour estimated on the best-fit straight line to the red sequence, $\sigma_c$ is the intrinsic scatter and $\kappa_l$ and $\kappa_h$ are numbers chosen after visual inspection of the colour-magnitude diagram as the best to bracket the red sequence. The modelling procedure and the estimates of the uncertainties on the parameters are all discussed in \cite{Cerulo_2016}.

The observed colour-magnitude diagrams of the HCS clusters are shown in Figure \ref{fig1}. In each plot are also shown the fit and boundaries of the red sequence (solid and dotted black lines, respectively) and the 90\% magnitude completeness limit (dashed diagonal line). The coloured symbols represent the galaxies that we classified for the analysis presented in this paper and are discussed in the following sections.

\begin{table*}
  \caption{The sample of the HAWK-I Cluster Survey (HCS) used in \protect\cite{Cerulo_2016}. Clusters are ordered by increasing redshift. The dark matter halo masses $M_{DM}$ in the fifth column from the left are taken from \protect\cite{Jee_2011}.}
  \begin{minipage}{14 cm}
  \begin{tabular}{|l|l|l|c|c|c|}
    \hline
     \multicolumn{1}{|c}{Cluster Name}  & \multicolumn{1}{c}{$\alpha$ (J2000)} & \multicolumn{1}{c}{$\delta$ (J2000)} & \multicolumn{1}{c}{Redshift} &  \multicolumn{1}{c}{$M_{DM}$} & \multicolumn{1}{c|}{Spectroscopically} \\
     \multicolumn{1}{|c}{}  & \multicolumn{1}{c}{} & \multicolumn{1}{c}{} & \multicolumn{1}{c}{} & \multicolumn{1}{c}{($10^{14} M_\odot$)} & \multicolumn{1}{c|}{Confirmed}   \\
     \multicolumn{1}{|c}{} & \multicolumn{1}{c}{} & \multicolumn{1}{c}{} & \multicolumn{1}{c}{} & \multicolumn{1}{c}{} & \multicolumn{1}{c|}{Members}  \\                  
    \hline
     \hline
     RX J0152.7-135 (RX0152) & 01:53:00 & -13:57:00 & 0.84 & $4.4^{+0.7}_{{-}{0.5}}$ & 130 \\
     RCS 2319.8+0038 (RCS2319) & 23:19:53.9 & +00:38:13 & 0.91 & $5.8^{+2.3}_{{-}1.6}$ & 30  \\
     XMMU J1229+0151 (XMM1229) & 12:29:28.8 & +01:51:34 & 0.98 & $5.3^{+1.7}_{{-}1.2}$  & 18   \\
     RCS 0220.9-0333 (RCS0220) & 02:20:55.7 & -03:33:19 & 1.03 & $4.8^{+1.8}_{{-}1.3}$ &  21    \\
     RCS 2345-3633 (RCS2345) & 23:45:27.3 & -36:32:50 & 1.04 & $2.4^{+1.1}_{{-}0.7}$  &  29   \\
     XMM J0223-0436 (XMMU0223) & 02:23:03.7 & -04:36:18 & 1.22 & $7.4^{+2.5}_{{-}1.8}$  & 20    \\
     RDCS J1252.9-2927 (RDCS1252) & 12:52:00 & -29:27:00 & 1.24 & $6.8^{+1.2}_{{-}1.0}$ &  42   \\
     XMMU J2235.3-2557 (XMMU2235) & 22:35:00 & -25:57:00 & 1.39 & $7.3^{+1.7}_{{-}1.4}$  & 25   \\
     XMM J2215-1738 (XMMXCS2215) & 22:15:58.5 & -17:38:02 & 1.45 & $4.3^{+3.0}_{{-}1.7}$ &  26  \\
  \hline
  \end{tabular}
\end{minipage}
\label{table1}
\end{table*}

\subsection{The low-redshift Comparison Samples}

In order to study the morphological evolution of red sequence galaxies, we built a comparison sample of morphologically classified galaxies from WINGS-SPE, the spectroscopic follow-up of the WIde-field Nearby Galaxy-cluster Survey (WINGS, \citealt{Fasano_2006}, \citealt{Cava_2009}).

Morphologies in this sample are presented in \cite{Fasano_2012}. Galaxies were classified on optical V-band images (\citealt{Varela_2009}), using a neural-network approach, in 18 types approximately corresponding to the Hubble T-types \citep{DeVauc_1959}. We converted the WINGS morphologies to our morphological scheme as illustrated in Table 5 of \cite{Cerulo_2014} and selected only the clusters with masses $M_{DM} > 5 \times 10^{14} M_\odot$, which, according to the predictions of simulations of structure formation in a $\Lambda CDM$ universe (e.g.\ \citealt{Fakhouri_2010}, \citealt{Chiang_2013}, \citealt{Correa_2015a, Correa_2015b, Correa_2015c}), is approximately the mass of the descendant of the lowest-mass cluster in the HCS sample (i.e.\ RCS 2345-3633 at $z=1.04$). We estimated the halo masses of the WINGS clusters using the velocity dispersions reported in Tables 1 and 2 of \cite{Cava_2009}. Our selection resulted in a subsample of 29 WINGS clusters which are likely descendant of the HCS clusters.

Following \cite{Cerulo_2014} and \cite{Cerulo_2016}, we fitted a straight line to the rest-frame $(B-V)_{AB}$ vs $V_{AB}$ red sequence of the composite WINGS sample considering only galaxies within $0.54 \times R_{200}$ from the centre of each WINGS cluster. Then we selected all the galaxies with $V_{AB, obs} < 18.0$ mag, corresponding to the 50\% spectroscopic completeness of WINGS-SPE. We note that $0.54 \times R_{200}$ in WINGS is about twice the spatial extent of $0.54 \times R_{200}$ at the redshifts of HCS and, for this reason, we also estimated morphological fractions in WINGS within a radius $0.27 \times R_{200}$.

In addition to the WINGS clusters we selected the clusters cl1232.5−1250 ($z=0.54$), cl1054.4-1146 ($z=0.70$), cl1054.7-1245 ($z=0.75$), and cl1216.8-1201 ($z=0.80$), from the ESO Distant Cluster Survey (EDisCS, \citealt{White_2005}), which have halo masses consistent with those expected from the likely descendants of the HCS clusters at these redshifts. We estimated the halo masses for these four clusters using the velocity dispersions reported in Table 1 of \cite{Poggianti_2006}.  We only used spectroscopically confirmed cluster members as defined in \cite{Halliday_2004} and \cite{Milvang_Jensen_2008} and took into account the incompleteness of the spectroscopic sample by associating a weight with each galaxy following the method outlined in Appendix A of \cite{Poggianti_2006}. 

We used the morphological classification published in \cite{Desai_2007} performed on HST/ACS images taken with the F814W filter. This morphological sample is composed of galaxies with magnitudes $I_{814, AUTO, Vega} < 23.0$ mag, where $I_{814, AUTO, Vega}$ is the {\ttfamily{SExtractor}} (\citealt{Bertin_1996}) {\ttfamily{AUTO}} magnitude measured in the Vega photometric system. This selection corresponds to galaxies with apparent AB magnitudes  $I_{814, AUTO, AB} < 23.4$. \cite{Desai_2007} visually classified galaxies in EDisCS following the T type scheme introduced in \cite{DeVauc_1959}, where a number between -5 and 11 is assigend to galaxies with morphologies ranging from elliptical to irregular. This classification was translated into the system introduced in \cite{Cerulo_2014} and adopted in the present paper in a similar fashion as done for the WINGS sample. 

We selected galaxies on the red sequence within 700 kpc from the centre of each cluster. This radius approximately corresponds to $0.54 \times R_{200}$ at the redshifts of HCS and $0.27 \times R_{200}$ at the redshifts of WINGS, thus allowing one to compare morphological fractions within the same spatial volume. At the redshifts of the EDisCS clusters the $I$ F814W band spans a rest-frame wavelength range corresponding to the B and V optical bands.

\subsection{Stellar Mass Estimate of Red Sequence Galaxies}

We used the software {\ttfamily{lephare}} \citep{Arnouts_1999, Ilbert_2006} to estimate stellar masses of red sequence galaxies in the HCS. {\ttfamily{lephare}} is based on a $\chi^2$ minimisation algorithm which fits the available photometry to a library of template spectral energy distributions (SED) that can be set in the input configuration file. The SED fitting was performed on each cluster photometric data-set, exploiting all the available photometry. The construction of the multiwavelength photometric catalogues is explained in \cite{Cerulo_2016}, and we briefly summarise it in this section.

We created multiband photometric catalogues for each cluster field by matching the PSF of each image to the broadest one. We estimated magnitudes within 1'' radius apertures in the PSF-matched images. As discussed in \cite{Cerulo_2014}, this aperture size, which corresponds to a physical projected radius that varies from 7.6 kpc at $z=0.84$ to 8.5 kpc at $z=1.46$, allows us to avoid the effects of colour gradients for bright galaxies and to include a large fraction of the least luminous galaxies. The photometric uncertainties were derived by estimating the sky noise in varying apertures randomly placed in regions of the images where no objects were detected by SExtractor.

We used a set of templates drawn from the \cite{Bruzual_2003} library with three different metallicities ($0.4 Z_\odot$, $Z_\odot$, $2.5 Z_\odot$), exponentially declining star formation rates, and \cite{Chabrier_2003} initial mass function (IMF), and we fitted them to the multiwavelength photometry available for each cluster. This set of templates was the same adopted by \cite{Delaye_2014}, who studied the stellar mass vs size relation in the HCS, allowing us to make a direct comparison with that work. We estimated again the stellar masses in XMM1229 using this set-up and the multiband sample of \cite{Cerulo_2014}. For each cluster field we considered all galaxies at the redshift of the cluster shown in Table \ref{table1}.

In order to estimate the uncertainties on stellar mass we followed the Monte Carlo approach outlined in \cite{Arnouts_1999}. We built simulated photometric catalogues perturbing the magnitudes of the galaxies in each band by a random amount $\varepsilon$ included in the range $-\delta m < \varepsilon < +\delta m$, where $\delta m$ is the photometric error. We generated 20 simulated objects for each real object in the catalogues and obtained simualted data-sets containing up to 54000 objects. We ran {\ttfamily{lephare}} on the simulated catalogues and evaluated the $1\sigma$ width of the simulated stellar mass distribution in two-dimensional bins of 1.0 mag in $z_{850}$ and 0.5 dex in stellar mass. The stellar mass uncertainties estimated with this Monte Carlo technique range between 0.02 dex and 0.25 dex. Althuough the least massive galaxies ($log(M_*/M_\odot) < 10.0$) tend to have larger uncertainties (0.10-0.25 dex), the most massive galaxies exhibit error values that span the range 0.02-0.20 dex. 

We note that the uncertainties estimated in this way only take into account the effect of the random photometric error. We did not experiment by changing the IMF or the template set. \cite{Ilbert_2010} show that different template sets can bias the stellar mass estimate by an amount of up to 0.15 dex. We did not correct our stellar mass estimate for such systematic offsets.

In this study, which is focused on red sequence galaxies, we did not use any dust correction in the SED fitting. We showed in \cite{Cerulo_2016} that the HCS red sequence is poorly contaminated by dusty star-forming galaxies.

We finally note that the photometric coverage for some of the clusters is relatively poor, comprising only 4 photometric bands (RX0152, RCS2319, RCS0220, RCS2345, XMMU0223, XMMXCS2215, see Table 2 in \citealt{Cerulo_2016}). However, in all the cases there are at least two photometric bands covering wavelengths redder than 4000 \AA\ at the redshift of each cluster, ensuring that the stellar mass of quiescent galaxies is not significantly underestimated.

For the galaxies in common with the \cite{Delaye_2014} sample we find a median difference $\Delta M = \log(M_*/M_\odot) - \log(M_{Delaye}/M_\odot) = -0.02 \pm 0.12$, where the error corresponds to the 68\% width of the $\Delta M$ distribution. This difference translates into a median ratio $(M_*/M_{Delaye}) \sim 0.95$. We attribute this residual difference to the different strategies adopted for photometry, as \cite{Delaye_2014} used unconvolved total ({\ttfamily{SExtractor}} {\ttfamily{MAG\_AUTO}}) magnitudes whereas we use PSF matched $1''$ radius aperture magnitudes, and to the different cosmologies adopted in the two works\footnote{\cite{Delaye_2014} adopted a $\Lambda CDM$ cosmology with $H_0=70$ km $\cdot$ s$^{-1} \cdot$ Mpc$^{-1}$, $\Omega_\Lambda=0.7$, and $\Omega_m = 0.3$}. Furthermore, the different numbers of photometric bands used for some of the clusters (i.e.\ RDCS1252, XMMU2235, XMM1229) are also likely to contribute to $\Delta M$.

In order to compare clusters at different redshifts we built a mass-complete sample and estimated the stellar mass limit following two methods. The first method, described in \cite{Pozzetti_2010}, involves estimating the mass of a galaxy, at a certain redshift, under the assumption that it is observed at the magnitude limit of the survey and that the stellar mass-to-light ratio of galaxies is constant at all luminosities. This leads, for a galaxy of stellar mass $M_*$, observed at magnitude $z_{850}$ in the ACS F850LP band, to the equation:
\begin{equation}\label{eq2}
\log(M_{lim}/M_\odot) = log(M_*/M_\odot) - 0.4(z_{lim} - z_{850})
\end{equation}
where $M_{lim}$ is the limiting mass and $z_{lim}$ the limiting magnitude of the morphological sample, i.e.\ $z_{lim} = 24.0$ mag (see Section 2.4). We computed $M_{lim}$ for the 20\% faintest galaxies in the morphological sample of each cluster and took the upper 95\% boundary of the $M_{lim}$ distribution as the mass completeness limit of our sample. By restricting the calculation of $M_{lim}$ to the 20\% faintest galaxies of the morphological sample, only galaxies with stellar mass-to-light ratios typical of the faint end of the morphological sample are considered, and the inclusion of the most massive galaxies is avoided. We applied Equation \ref{eq2} to the spectroscopically confirmed members of each HCS cluster and obtained $\log(M_{lim}/M_\odot) = 11.1$ at $z=1.46$, the redshift of XMMXCS2215, which is the most distant cluster in the HCS.

The second method (see e.g.: \citealt{Bundy_2005}, \citealt{Vulcani_2011} and \citealt{Delaye_2014}) consists in measuring the stellar mass of a hypothetical object having a spectral energy distribution equal to a model synthetic SED with solar metallicity, formation redshift $z_f=5.0$, \cite{Chabrier_2003} IMF, and exponentially declining star formation rate with e-folding time $\tau=0.5$ Gyr. Assuming that the object has $z_{850} = 24.0$ mag, and that the galaxy stellar mass-to-light ratio is constant at all luminosities, we obtain from Equation \ref{eq2} $\log(M_{lim}/M_\odot) = 10.7$ at $z=1.46$. 

This value is 0.3 dex lower than that obtained with the \cite{Pozzetti_2010} method. Furthermore, our estimate of the stellar mass completeness limit with the \cite{Pozzetti_2010} method is also 0.3 dex higher than the stellar mass limit obtained with the same method in \cite{Delaye_2014}. However, we note that the spectroscopic red sequence sample used in the estimation of the stellar mass completeness limit is small, and the 20\% faintest galaxies in XMMXCS2215 only contain 10 objects. With the spectroscopic sample being so small, a difference in the stellar mass measurement between this work and \cite{Delaye_2014} of the order reported above in this section may induce the 0.3 dex difference  that we find in the stellar mass limit. In the following we decide to adopt the stellar mass limit $\log(M_*/M_\odot) = 10.9$, which corresponds to the mean of the limits obtained with the \cite{Pozzetti_2010} method and with the passive evolution of the synthetic stellar population. We stress that applying a limit closer to the 10.7-10.8 dex of \cite{Delaye_2014} does not change the results of our analysis (see Figure \ref{fig5}).

Stellar masses in WINGS-SPE were estimated by \cite{Fritz_2011}, who adopted a model fitting approach in which all the available optical and near-infrared photometry, and the spectra, were used to estimate stellar masses and stellar ages of cluster members. A \cite{Salpeter_1955} IMF was assumed to fit simple stellar population (SSP) models to the WINGS photometry and spectra. The stellar mass of the WINGS galaxies is defined as the total mass of stars, both in their nuclear-burning phase and in the remnants (white dwarfs, neutron stars and black holes). These values of the stellar masses were corrected to account for radial colour gradients by adding an empirical correction term provided in the WINGS catalogues and defined in Equation 3 of \cite{Fritz_2011}. The average error on the estimates of the total stellar masses in WINGS is 0.2 dex \citep{Moretti_2014}. We applied an offset of -0.25 dex to account for the difference between stellar mass estimates with a \cite{Salpeter_1955} IMF and a \cite{Chabrier_2003} IMF (see \citealt{Bernardi_2010}). In this way we mitigated the effects of systematic errors resulting from assuming two different IMFs in the estimation of stellar masses in HCS and WINGS.

We did not estimate stellar masses for the EDisCS galaxies as the photometry available in the public catalogues, including only the B, V and I or V, R and I bands, was not sufficient to obtain reliable stellar masses through SED fitting.

\subsection{Morphology of HCS Galaxies}

\subsubsection{Classification Procedure}

The procedure adopted for the morphological classification of the HCS red sequence galaxies is analogous to that described in \cite{Cerulo_2014} for the cluster XMM1229 at $z=0.98$. The HST/ACS F850LP images, which are the deepest available in the HCS sample, were used to perform the morphological classification in all the clusters. The $z_{850}$ band corresponds to the SDSS $g$ band in the redshift range $0.8 < z < 1.24$ and to the SDSS $u$ band at redshifts $1.3 < z < 1.5$. Since all our galaxies are on the red sequence, the effect of morphological k-correction is not significant, and we can directly compare with the morphologies of red sequence galaxies in EDisCS and WINGS.

We selected all red sequence galaxies within $0.54 \times R_{200}$ from the cluster centre with total ({\ttfamily{SExtractor}} {\ttfamily{MAG\_AUTO}}) magnitudes $z_{850} < 24.0$ mag, the limit down to which we were able to distinguish morphological features (\citealt{Postman_2005, Cerulo_2014, Delaye_2014}). The selection resulted in a subsample of 428 galaxies that we will refer to as the {\itshape{morphological sample}} in this paper. The morphological sample also includes the XMM1229 members classified in \cite{Cerulo_2014}. As shown in Figure \ref{fig1}, the limiting magnitude $z_{850} = 24.0$ mag of the morphological sample does not allow us to study the morphology of galaxies down to the faint end of the red sequence in all the clusters. For example, we cannot study the morphology of galaxies at the faint end of the red sequence in the clusters RCS2319, RDCS1252 and XMMU2235, while in the cluster XMMXCS2215 only galaxies which are bluer than the best-fit line to the observed red sequence fall in the morphological sample. We will discuss the implications of this limit in Section 4.

We divided the morphological sample into five types which were based on the apparent bulge-to-total light ratio, namely, ellipticals ({\itshape{pure bulges}}, E), bulge-dominated (BD), which correspond to S0/S0a galaxies, early-type disc-dominated (EDD), corresponding to Hubble types in the range Sa-Sbc, late-type disc-dominated (LDD), corresponding to Hubble types in the range Sc-Scd, and irregular galaxies (Irr). Since only 30 galaxies (7\%) of the morphological sample have late-type disc or irregular morphologies, we decided to merge these two classes into one family of LDD+Irr galaxies (blue squares in Figure \ref{fig1}). Therefore, as previously done in the study of the morphological properties of XMM1229, in the following we will present the results of the investigation of the morphological evolution of the four classes of E, BD, EDD and LDD+Irr galaxies. Since most of the S0 galaxies fall into the class of bulge-dominated, we will use these two terms interchangeably.

We adopted a coupled visual and automated classification procedure whereby each galaxy was visually inspected by three independent classifiers (i.e.\ P.C., W.J.C and C.L.) and a fourth independent classification was obtained by running the {\ttfamily{galSVM}} software \citep{Huertas_2008, Huertas_2011} on the F850LP images. {\ttfamily{galSVM}} is an IDL package based on Support Vector Machines (SVM), which are machine learning algorithms particularly suited for the solution of problems of classification in large samples. The software also implements an {\itshape{a posteriori}} estimate of the probability for each galaxy to be of a certain morphological type. This allows the uncertainty in the classification to be quantified. 

The morphological type was defined as the mode of the four classifications. In cases in which two classifiers agreed on one type and the other two on a different type, the type corresponding to the earliest morphological class was assigned to the galaxy. Thus, if two classifiers classified a galaxy as elliptical and the other two as S0, the galaxy was assigned to the morphological class of the ellipticals. In fact, with the cosmological surface brightness dimming, the disc of the faintest galaxies falls below the level of the sky noise and, therefore, the chances of \textcolor{black}{not} detecting a disc in the visual classification are high. In this way the halo of an elliptical galaxy can be confused with a low-inclination disc. We will discuss this in more detail, together with the overall reliability of the morphological classification, in the next section. 
Galaxies for which each classifier assigned a different type were not assigned to any class. Only 7 objects (2\%) are unclassified in the HCS morphological sample.

The ACS F850LP postage stamp images of the galaxies classified in the cluster RDCS1252 are shown as example of our morphological classification in Appendix A in Figure A1.

\begin{table*}
  \caption{Mean values and standard errors of concentration, asymmetry, Gini coefficient and $M_{20}$. The estimates for each morphological type and for early- and late-type galaxies are all shown.}
  \hspace{-0.5cm}
  \begin{tabular}{|c|c|c|c|c|}
    \hline
     \multicolumn{1}{|c}{Morphological}  & \multicolumn{1}{c}{Concentration} &  \multicolumn{1}{c}{Asymmetry}  & \multicolumn{1}{c}{Gini Coefficient} & \multicolumn{1}{c|}{$M_{20} \pm \delta M_{20}$} \\
     \multicolumn{1}{|c}{Type}  & \multicolumn{1}{c}{($C_{Con} \pm \delta C_{Con}$)} & \multicolumn{1}{c}{$A \pm \delta A$}  & \multicolumn{1}{c}{($G \pm \delta G$)} & \multicolumn{1}{c|}{ } \\
     \hline
      \hline
                            &                       &                        &                      &                      \\
      Elliptical            & $2.81 \pm 0.02$       &  $0.0729 \pm 0.005$    & $0.511 \pm 0.003$    &  $-1.620 \pm 0.016$   \\
                            &                       &                        &                      &                        \\
      Bulge-Dominated       & $2.78 \pm 0.02$       &  $0.073 \pm 0.004$     & $0.538 \pm 0.004$    &  $-1.617 \pm 0.018$    \\
                            &                       &                        &                      &                        \\
                            &                       &                        &                      &                        \\
      Early-type            & $2.47 \pm 0.04$       & $0.076 \pm 0.008$      & $0.473 \pm 0.007$    &  $-1.56 \pm 0.03$       \\
      Disc-Dominated        &                       &                        &                      &                         \\
                            &                       &                        &                      &                          \\
      Late-type             & $2.13 \pm 0.04$       & $0.08 \pm 0.02$        & $0.414 \pm 0.008$    &  $-1.33 \pm 0.04$        \\
      Disc-Dominated + Irr  &                       &                        &                      &                           \\
                            &                       &                        &                      &                            \\
      \hline
                            &                       &                        &                      &                            \\
      early-type galaxies   & $2.801 \pm 0.016$     & $0.073 \pm 0.004$      & $0.520 \pm 0.003$    &  $-1.619 \pm 0.012$        \\
      ( Elliptical + S0 )   &                       &                        &                      &                            \\
                            &                       &                        &                      &                             \\
      late-type galaxies   & $2.35 \pm 0.03$       & $0.078 \pm 0.009$      & $0.452 \pm 0.006$    &  $-1.48 \pm 0.02$           \\
      ( disc-dominated + Irr )   &                       &                        &                      &                             \\
 \hline
  \end{tabular}
  \label{table2}
\end{table*}

\subsubsection{Testing Morphology I: Internal Comparison}

To assess the reliability of our morphological classification, we first compare the four independent classifications and then the visual classification with that from {\ttfamily{galSVM}}.

There was full agreement among the four classifiers on only 77 galaxies, which correspond to 18\% of the morphological sample, while the three human classifiers agreed on assigning the same type to 173 galaxies, corresponding to 40\% of the morphological sample. These fractions are an improvement on the agreement found in the classification of red sequence galaxies in XMM1229 (8\% for the agreement of the four classifiers and 14\% for the agreement of the human classifiers), but are still low and underline the difficulties in the morphological classification of distant galaxies. The agreement between classifiers improves if the four morphological classes are merged into broad classes of early- (i.e.\ elliptical+S0) and late-type (i.e.\ disc-dominated and irregular) galaxies. In this case the four classifiers agree in assigning the same early or late type to 57\% of the sample, while the three human classifiers agree in assigning the same type to 70\% of the sample. Although there is still a considerable disagreement in the classification of early- and late-type galaxies (43\% between all classifiers and 30\% between human classifiers), we see that the agreement in the classification, when the distinction between elliptical and S0 galaxies is removed, improves by a factor of 3 (2 if only visual classification is taken into account). This suggests that the main source of error in our classification lies in the separation between elliptical and S0 galaxies.

Such a conclusion is not surprising since, as it is well known in the literature (e.g.\ \citealt{Abraham_2007}, \citealt{Desai_2007}, \citealt{Mei_2009}, \citealt{Huertas_2008, Huertas_2011}), face-on S0 galaxies can be easily misclassified as elliptical galaxies. This difficulty is exacerbated in distant galaxies, where the cosmological surface brightness dimming, which has a $(1+z)^4$ redshift dependence, causes the lowest-surface-brightness features, such as discs and spiral arms, to become fainter than the sky background. As a result, in the faintest S0 or disc-dominated galaxies with low inclination, only the bulge is visible, and they are classified as ellipticals.

Interestingly, we note that when comparing the visual classification, defined as the mode of the three individual visual classifications, and the {\ttfamily{galSVM}} classification, 19\% of the objects with early-type disc-dominated visual morphology are classified as elliptical galaxies by {\ttfamily{galSVM}}. This mismatch is explained by the fact that the automated classification, which relies upon the measurement of morphological coefficients based on the light distribution of galaxies, is even less efficient in detecting low-surface-brightness features in the images. We also note that 33\% of the galaxies with visual S0 type were classified by {\ttfamily{galSVM}} as early-type disc-dominated. This mismatch certainly underlines the difficulties inherent in the distinction between two galaxy types based only on bulge-to-total ratio.

\begin{figure*}
	\subfloat[]{\label{fig:a}}\includegraphics[width=0.4\textwidth, trim=0.0cm 2.5cm 0.0cm 0.0cm, clip, page=1]{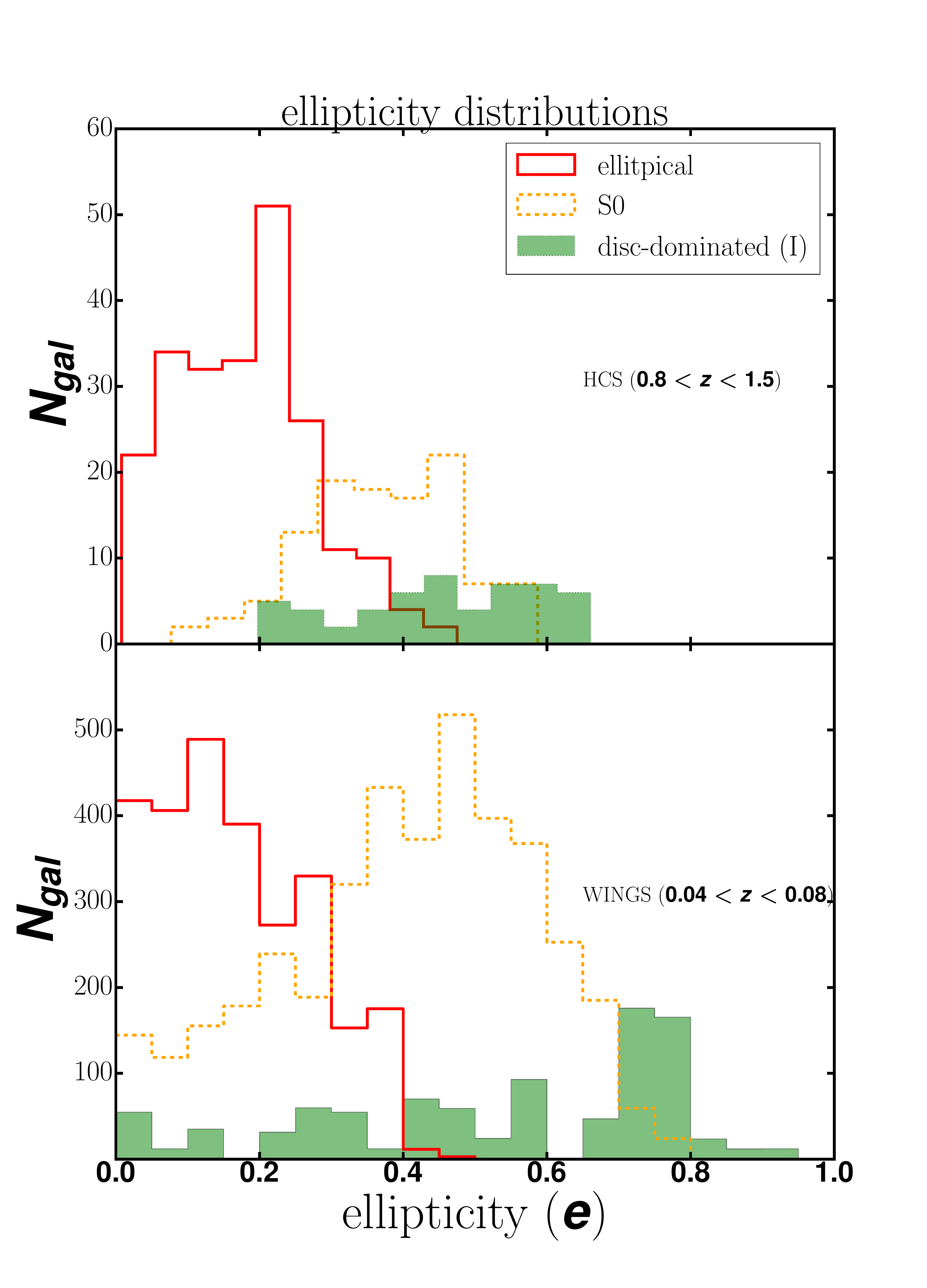}
        \hspace{1mm}
        \subfloat[]{\label{fig:a}}\includegraphics[width=0.4\textwidth, trim=0.0cm 2.5cm 0.0cm 0.0cm, clip, page=1]{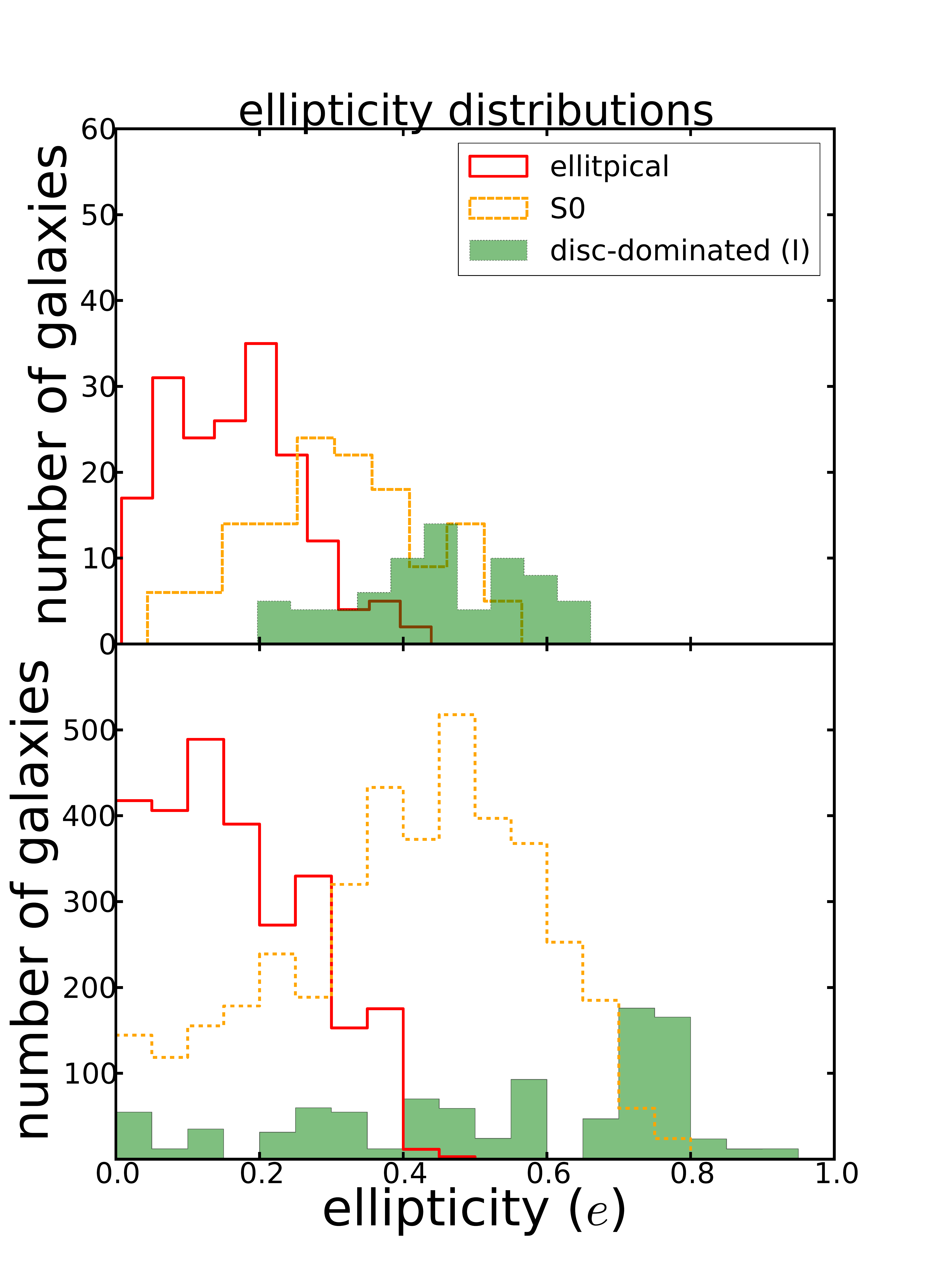}
	\caption{(a): Distribution of the ellipticity for galaxies classified as ellipticals, bulge-dominated and early-type disc-dominated in the HCS (top) and WINGS (bottom) samples. (b): the same as in (a) but with the alternative classification discussed in \S 2.4.3. Galaxies for which 50\% of the classifiers agreed on one type and the other 50\% on a different type are now assigned to the latest-type class of the two. If 2 classifiers classified the galaxy as elliptical and 2 classified it as S0, the galaxy was assigned to the class of the S0s. In this way part of the population of S0 galaxies with low ellipticity (nearly face-on, $e<0.2$) is recovered. With both classification strategies we find few S0 galaxies with low ellipticity in the HCS sample. Although this result suggests that we may be missing face-on S0 galaxies, previous works found that the fraction of low-ellipticity S0 galaxies decreases with redshift in clusters of galaxies (see Section 2.4.3.)}
\label{fig2}
\end{figure*}

\begin{figure*}
	\includegraphics[width=0.8\textwidth, trim=0.0cm 15.0cm 0.0cm 6.0cm, clip, page=2]{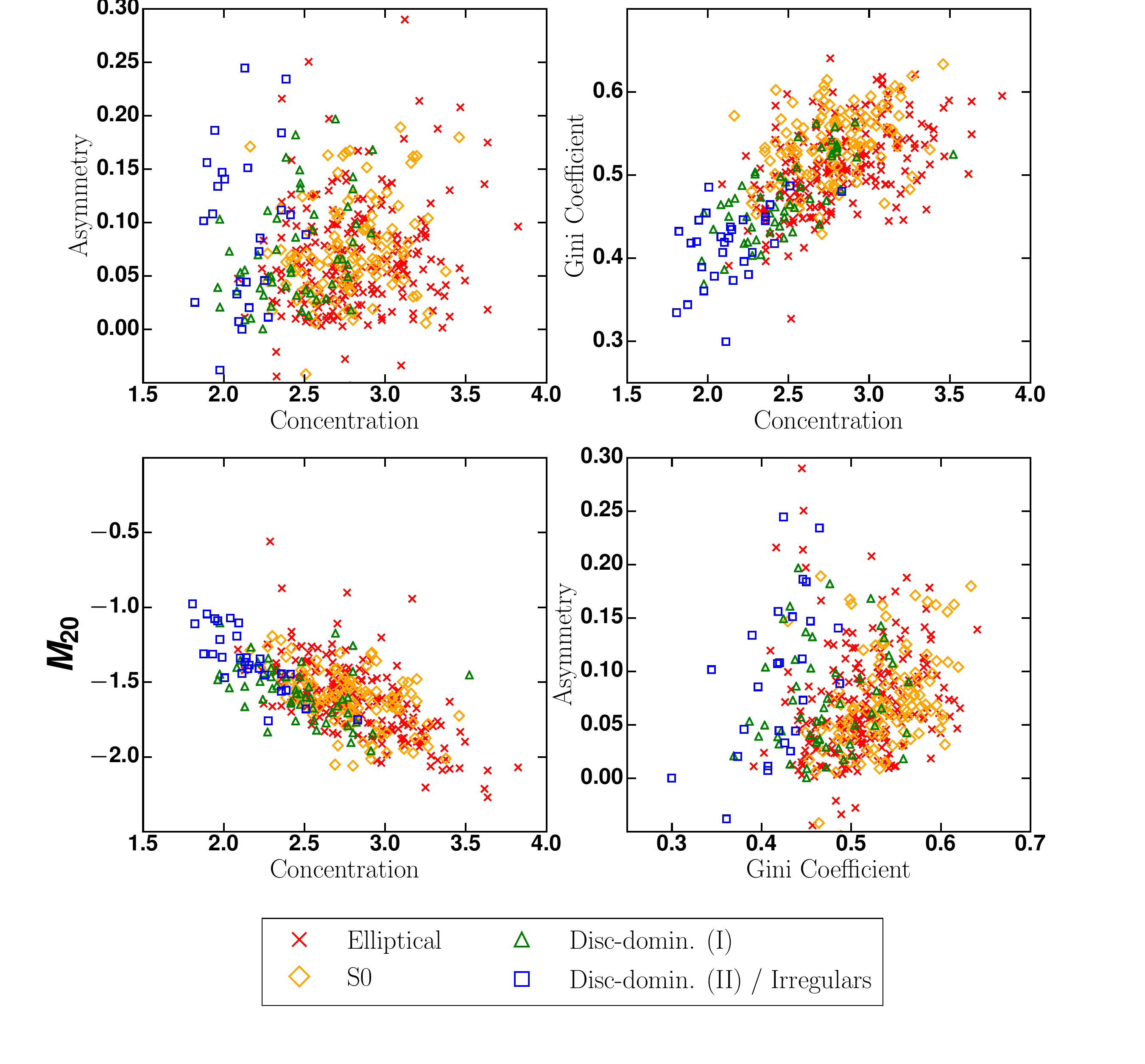}
	\caption{Morphological parameters of HCS red sequence galaxies as estimated by {\ttfamily{galSVM}}. {\itshape{Top-left}}: asymmetry vs concentration, {\itshape{top-right}}: Gini coefficient vs concentration, {\itshape{bottom-left}}: $M_{20}$ vs concentration, {\itshape{bottom-right}}: asymmetry vs Gini coefficient. A combination of concentration, Gini coefficient and $M_{20}$ is effective in dividing the sample between early- and late-type galaxies. Disc-dominated and irregular galaxies are less concentrated, have lower Gini coefficients and higher $M_{20}$ values, indicating the presence of a higher level of substructure with respect to elliptical and S0 galaxies.}
\label{fig3}
\end{figure*}

\subsubsection{Testing Morphology II: The Elliptical vs S0 Separation}

In order to test the reliability of the separation between elliptical and S0 galaxies we study the distributions of 6 morphological parameters, namely ellipticity, concentration, asymmetry, Gini coefficient, second order moment of the light distribution, and S\'{e}rsic index. 

The ellipticity $e$ of galaxies is a proxy for both inclination and bulge-to-total light ratio \citep{Binney_Merrifield_1998} and, in particular, the ellipticity is directly proportional to the inclination of galaxies, so that face-on galaxies have $e\sim0$ and edge-on galaxies have $e \sim 1$. Figure \ref{fig2} shows the distribution of the ellipticity of galaxies classified as ellipticals (red), bulge-dominated (orange), and early-type disc-dominated, in both the HCS (upper panels) and WINGS samples (bottom panels). The top-left panel shows the distributions for the classification adopted in this work. In the top-right panel we change the criterion for the attribution of the morphological class of those HCS galaxies for which two classifiers agreed on a type and the other two agreed on a different type and, rather than assigning the earliest morphological type between the two, we assign the latest. Thus, if it happened that two classifiers assigned a galaxy to the class of the ellipticals and the other two to the class of the S0s, the galaxy was classified as S0. In this way it is possible to evaluate the contamination of the elliptical and S0 samples due to the arbitrary assignment of a galaxy, for which 50\% of the classifiers agreed on a certain type and 50\% on a different type, to a particular morphological class. In this test we only consider the first three morphological classes since late-type discs and irregular galaxies represent together only 7\% of the morphological sample. The histograms of the WINGS galaxies are corrected for the incompleteness of the spectroscopic sample.

The median values of the ellipticities, and the boundary of the 68\% confidence intervals, are shown in Table \ref{table3}. We note that the median values of the HCS ellipticities, for each morphological type, are consistent within $1\sigma$ with the median values measured in WINGS. However, the WINGS sample exhibits a population of low-ellipticity S0 and early-type disc-dominated galaxies, at $e<0.1$, which we do not observe in the HCS sample even if we adopt the alternative classification scheme (top-right panel). This reflects the difficulty in detecting face-on discs in our morphological classification due to the cosmological surface brightness dimming. This effect is especially severe for passive galaxies where the absence of star-formation makes discs fainter in the U, B and V bands, which are the regions of the galaxy spectrum probed by the F850LP images at the redshifts of the HCS clusters.

\cite{Vulcani_2011b}, using the visual morphological classification of \cite{Desai_2007} in EDisCS, found that only 5\% of the S0 galaxies at $0.4 < z < 0.8$ have $e<0.1$, suggesting that low-ellipticity lenticular galaxies become rarer towards high redshifts. Interestingly, \cite{Vulcani_2011b} also show that the number of low-ellipticity S0s is higher in nearby clusters. The reliability of the EDisCS classification was tested with galaxy light profiles, and the chance of misclassifications, particularly between elliptical and S0 galaxies, was shown to be low \citep{Desai_2007}. Given the rarity of low-ellipticity S0 galaxies in distant clusters, the contamination of the elliptical subsample from face-on S0s in HCS should be less than 10\% and thus it should not influence the conclusions of our analysis. However, we notice that \cite{Aguerri_2005} detected a population of dwarf (i.e.: $-18 < M_B < -16$) S0 galaxies in the Coma cluster that were visually and photometrically indistinguishable from elliptical galaxies at the same luminosities. Although these galaxies could belong to the same visual class and have colours similar to the dwarf ellipticals, their light profiles were better fitted by a two-component profile with a S\'{e}rsic-like bulge and an exponential disc. We cannot probe the luminosity range of \cite{Aguerri_2005} in none of the three samples used in this work, yet we do not exclude that at brighter luminosities galaxies that are visually classified as ellipticals may be better fitted by two-component profiles.

Interestingly, we note that the WINGS elliptical galaxies contain more low-ellipticity systems than the HCS. This property of the WINGS ellipticals is also mentioned in \cite{Vulcani_2011b}, who explained the existence of these galaxies as the result of galaxy mergers (see  \citealt{Holden_2009} for a different conclusion). We will not address this problem in the present work and here we only state that a two-sample Kolmogorov-Smirnov (KS) test returns a probability $P_{KS} \sim 0.02$ that the ellipticity distributions of WINGS and HCS elliptical galaxies were drawn from the same parent distribution. This implies that the distributions at $z=0$ and $z=1$ are different for cluster ellipticals and that elliptical galaxies tend to be rounder in WINGS with respect to HCS.

We also note that WINGS hosts a population of high-ellipticity disc-dominated and S0 galaxies, at $e>0.6$, which is not observed in HCS. This can be a result of the smaller angular sizes of distant galaxies, with respect to nearby galaxies, which approach the width of the PSF of the instrument. As a consequence, due to the convolution of their intrinsic light distributions with the PSF in the ACS F850LP filter, these objects appear rounder.

As a further test for the reliability of the elliptical vs S0 separation in HCS we compare four of the morphological parameters estimated by {\ttfamily{galSVM}}, namely concentration, asymmetry, Gini coefficient, and second order moment of the galaxy light distribution $M_{20}$. These particular coefficients provide information on the amount of concentration and level of homogeneity of the light distribution in galaxies. The planes formed by concentration and asymmetry (top-left panel), concentration and Gini coefficient (top-right panel), concentration and $M_{20}$ (bottom-left panel), and Gini coefficient and asymmetry (bottom-right panel) are shown in Figure \ref{fig3}. The mean values and standard errors of the four coefficients are shown in Table \ref{table2}. \cite{Huertas_2008} show that {\ttfamily{galSVM }} is highly efficient in separating early- and late-type galaxies using Concentration and Asymmetry; however, indices such as the Gini Coefficient and $M_{20}$ allow {\ttfamily{galSVM}} to distinguish between galaxies with or without substructures such as spiral arms, star-forming regions, and bars (see e.g.\ \citealt{Conselice_2000}, \citealt{Conselice_2003}, \citealt{Lotz_2004}, \citealt{Scarlata_2007a}).

We find that the Gini coefficient is particularly effective in separating elliptical from S0 galaxies. In fact, there is a $5\sigma$ difference between the mean values of the Gini coefficient for elliptical and S0 galaxies, the latter having a higher mean Gini coefficient. We also find that a two-sample KS test between the distributions of the Gini coefficient of elliptical and S0 galaxies returns $P_{KS} < 0.1$\%, rejecting the null-hypothesis that the two distributions are drawn from the same parent distribution. The discrepancy between the Gini coefficients of elliptical and S0 galaxies can be attributed to the presence of the disc in S0 galaxies, which results in a more \textcolor{black}{inhomogeneous} light distribution. The mean values and the uncertainties shown in Table \ref{table2} confirm that the 4 morphological coefficients considered in this section are effective, with the only exception of asymmetry, in separating between early- and late-type galaxies. Interestingly, as shown in the asymmetry vs Gini coefficient plane, disc-dominated galaxies are found at values of Gini Coefficient less than 0.5, in broad agreement with \cite{Meyers_2012}, who used these two parameters to select galaxies with different morphologies in the HST Cluster Supernova Survey. In all four panels of Figure \ref{fig3} it can be seen that early (red and orange symbols) and late-type galaxies (green and blue symbols) occupy two different regions of the planes formed by the morphological coefficients. Disc-dominated and irregular galaxies are less concentrated, have lower Gini coefficients, and higher $M_{20}$ values, indicating the presence of a higher level of substructure with respect to elliptical and S0 galaxies.

\begin{figure}
        \centering
	\includegraphics[width=0.4\textwidth, trim=0.0cm 5.0cm 0.0cm 40.0cm, clip]{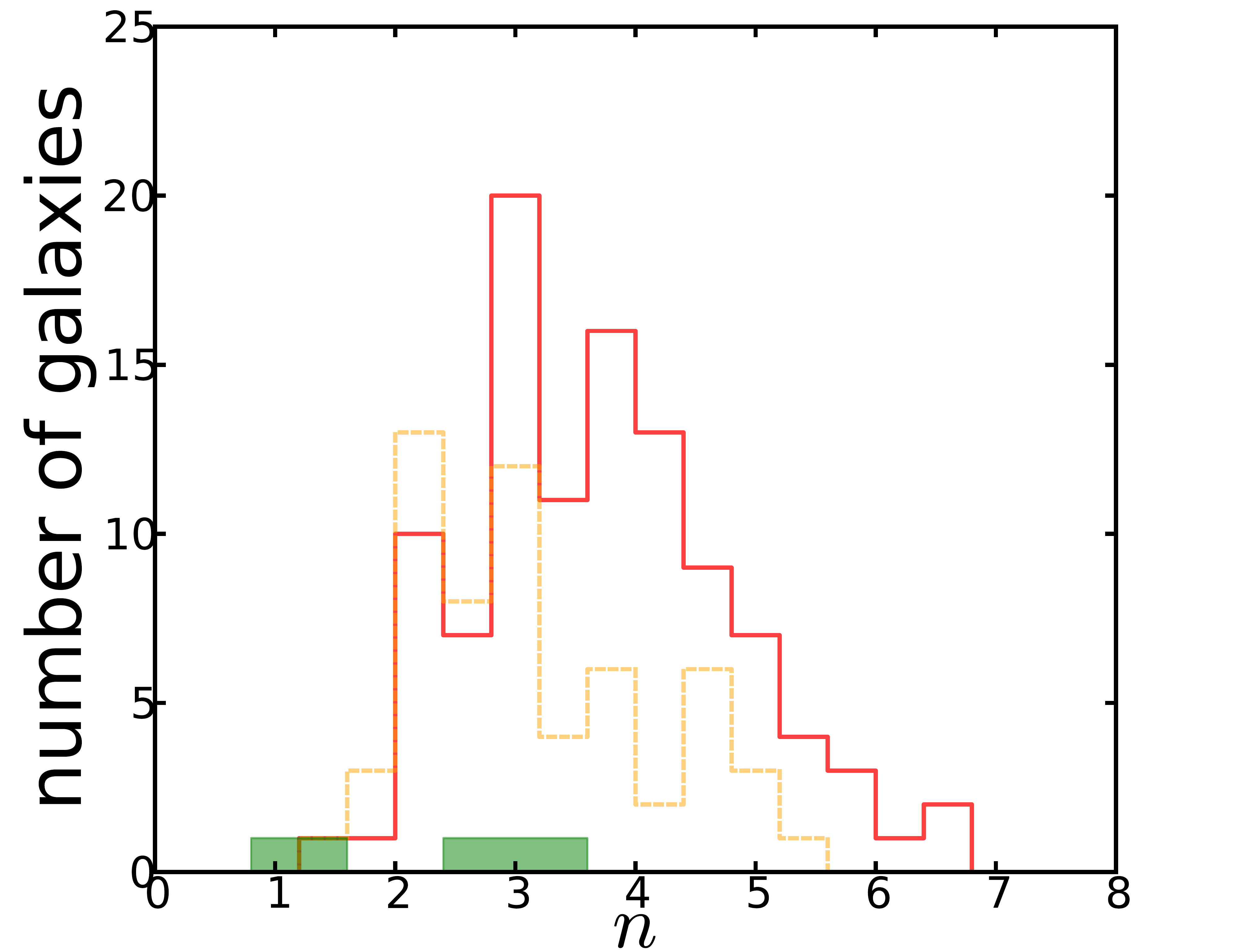}
	\caption{Distributions of the S\'{e}rsic index values, $n$, of elliptical (red histogram), bulge-dominated (orange histogram) and early-type disc-dominated (green filled histogram) galaxies in common with the sample of \protect\cite{Delaye_2014}. The distributions for elliptical and S0 galaxies are statistically different ($P_{KS} = 0.003$) and the mean values of $n$ are different at a $3.5 \sigma$ level (see \S 2.4.3).}
\label{fig4}
\end{figure}

Figure \ref{fig4} shows the distributions of the S\'{e}rsic index $n$ of the HCS galaxies from the measurements of \cite{Delaye_2014}\footnote{The catalogue with the structural parameters and stellar masses of HCS early-type galaxies is available with the electronic version of that paper.} for each morphological type. The mean and standard errors for the S\'{e}rsic indices of elliptical ($n_E$) and S0 ($n_{S0}$) galaxies (red and orange histograms, respectively) are: $\bar{n}_{E} = 3.72 \pm 0.10$ and $\bar{n}_{S0} = 3.15 \pm 0.13$. The difference between these two values is significant at a $3.5\sigma$ level. We can see that S0 galaxies tend to reside in the low-$n$ region of the distribution of $n_{E}$, suggesting that S0 galaxies tend to have lower values of the S\'{e}rsic index compared to elliptical galaxies. A two-sample KS test returns $P_{KS} = 0.003$ indicating that the two distributions are statistically different. This suggests that elliptical and S0 galaxies in our sample have different structural properties and that S0 galaxies, as expected, are less concentrated than elliptical galaxies.

The comparisons discussed in this section show that our morphological classification is reliable in separating elliptical from S0 galaxies, at least down to $z_{850} = 23.0$ mag. At fainter magnitudes the effects of the cosmological surface brightness dimming become severe and faint face-on S0s or disc-dominated galaxies can erroneously be classified as ellipticals. However, given the low number of high-redshift face-on S0 and disc-dominated galaxies observed in previous works (e.g.\ \citealt{Vulcani_2011b}), the contamination of the sample of red sequence ellipticals in HCS should not influence the conclusions of our analysis.

\subsubsection{Testing Morphology III: Comparison with the Literature}

In \cite{Cerulo_2014} we compared the morphological classification of XMM1229 with \cite{Santos_2009} and \cite{Delaye_2014} finding that our classification agreed with the visual classification of \cite{Santos_2009} for 9 of the 15 galaxies in common (13/15 if only the early- vs late-type separation was taken into account), and that 38/46 galaxies were classified as early-type both by us and by \cite{Delaye_2014}.

We can now repeat the same exercise with the entire morphological sample and compare with the morphological classifications published for other HCS clusters. \cite{Delaye_2014} published a catalogue of red sequence early-type galaxies classified with {\ttfamily{galSVM}}. Of the galaxies in common with our morphological sample, only 5\% were not classified as early-type (i.e.\ elliptical or S0) by us.

\cite{Hilton_2009} visually classified galaxies within 0.75 Mpc of the centre of the cluster XMMXCS2215, at $z=1.46$, using ACS F850LP images as in this work. Of the 16 galaxies in common with our sample, 13 were assigned the same types by \cite{Hilton_2009} and by us and, if we consider only the early- vs late-type separation, the two classifications agree in all the cases.

\cite{Blakeslee_2006} used the morphological classification of \cite{Postman_2005} for the cluster RX0152. After converting to our morphological scheme, we find that, of the 28 galaxies in common in the two samples, 18 (64\%) were assigned the same type in the two works. However, if we limit ourselves to the early- vs late-type subdivision, the classifications agree for 25 galaxies (i.e.\ 89\% of the common sample).

We conclude that our morphological classification of the HCS red sequence galaxies is robust also when compared to other classifications of the same galaxies published in the literature. Hence, it constitutes a reliable representation of the morphological content of the red sequence in clusters at redshift $0.8<z<1.5$.

\subsubsection{Testing Morphology IV: Sensitivity to rest-frame Wavelength}

As stated in Section 2.4.1, the ACS F850LP filter spans rest-frame wavelengths corresponding to the SDSS $u$ and $g$ bands at the redshifts of the HCS clusters. The emission of the old stellar populations typical of red sequence galaxies is low in the blue and near ultraviolet (NUV), leading to the potential loss of low surface brightness features, such as discs, in the images used for classification.

In order to test the effects of classifying red sequence galaxies at blue/NUV wavelengths we ran the following test. We selected 100 random red sequence galaxies, spanning the same redshift range of WINGS, from the spectroscopic Sloan Digital Sky Survey (SDSS, \citealt{York_2000}) database and one of the classifiers (P.C.) classified them visually in $u$- and $r$- band postage stamp images. The reason of choosing SDSS images is that their spatial resolution in kpc is similar to that of the F850LP images at the redshifts of the HCS clusters. The $r$ band is more sensitive to old stellar populations and so the low surface brightness components of red sequence galaxies should be more easily detectable in the images. The results of our test are summarised in Table \ref{table_extra}.

The comparison between the visually classified galaxies in the SDSS $u$- and $r$-band images shows that in the 69\% of the cases galaxies are assigned the same type in the two bands. 84\% of the galaxies classified as ellipticals in the $u$ band are also classified as ellipticals in the $r$ band, with a 10\% of them classified as S0s and 6\% as early-type disc-dominated. 56\% of galaxies classified as S0s in the $u$ band were also assigned a S0 type in the $r$ band, with a 33\% of them assigned to the elliptical class and 11\% to the early-type disc-dominated class. The classification of early-type disc-dominated galaxies presents a better agreement between the two photometric bands. Indeed 75\% of the galaxies that were classified as early-type disc-dominated in the $u$ band were also assigned the same type in the $r$ band, while in 25\% of the cases they were assigned to the S0 class. The class of the late-type disc-dominated+irregular galaxies shows the worst agreement between the two photometric bands. In only 18\% of the cases galaxies classified as irregulars or late-type disc-dominated in the $u$ band were also assigned to the same class in the $r$ band. In 41\% of the cases the galaxies in this class were assigned to the early-type disc-dominated class, while 24\% of them were classified as S0s and 18\% of them as ellipticals when seen in the $r$-band images.

This test shows that the classification of elliptical and early-type disc-dominated galaxies at blue/NUV wavelengths is robust, while the classifications of S0 and late-type disc-dominated galaxies are more affected by the wavelength difference. Interestingly, the galaxies that are classified as S0s in the $r$ band but not in the $u$ band (not shown in Table  \ref{table_extra}), are equally distributed between the elliptical and early-type disc-dominated classes. We stress, however, that we only classify 9 galaxies as S0s in the $u$ band and 19 in the $r$ band. With such numbers being small, we refrain from any statistically significant conclusion and only state that the use of blue/NUV images for visual morphological classification may affect the classification of S0 galaxies but without a particular preference for these to be classified as ellipticals.

\begin{table*}
  \caption{Dependence of visual morphological classification on wavelength. Each column shows, for a given morphological type in the SDSS $u$ band, the percentage of galaxies for each morphological class when they were classified in the $r$ band images (See Section 2.4.5).}
  \begin{minipage}[c]{0.8\textwidth}
    \centering
  \begin{tabular}{|c|c|c|c|c|}
     \hline
     \multicolumn{1}{|c}{}  & \multicolumn{1}{|c}{Elliptical}  & \multicolumn{1}{c}{Bulge-Dominated} & \multicolumn{1}{c}{Early-type} & \multicolumn{1}{c}{Late-type } \\
     \multicolumn{1}{|c}{}  & \multicolumn{1}{|c}{}  & \multicolumn{1}{c}{} & \multicolumn{1}{c}{Disc-Dominated} & \multicolumn{1}{c}{Disc-Dominated} \\
     \hline
      \hline
                            &           &           &           &          \\
      Elliptical            & $84$\%    & $33$\%    & $0$\%     & $18$\%    \\
                            &           &           &           &            \\
      Bulge-Dominated       & $10$\%    & $56$\%    & $25$\%    & $24$\%     \\
                            &           &           &           &            \\
                            &           &           &           &            \\
      Early-type            & $6$\%     & $11$\%    & $75$\%    & $41$\%      \\
      Disc-Dominated        &           &           &           &             \\
                            &           &           &           &              \\
      Late-type             & $0$\%     & $0$\%     & $0$\%     & $18$\%       \\
      Disc-Dominated + Irr  &           &           &           &               \\
                            &           &           &           &               \\
  \hline
  \end{tabular}
  \end{minipage}
  \label{table_extra}
\end{table*}

\section{Results}

The present section discusses the measurements of the morphological fractions and the statistical estimate of field contamination in each HCS cluster sample. 

\subsection{Morphological Fractions}

The number of spectroscopically confirmed cluster members in the HCS varies from cluster to cluster, spanning the range 18 (XMM1229) to 130 (RX0152). The spectroscopic sample does not probe the red sequence down to faint luminosities, close to the $z_{850} = 24.0$ mag limit of the morphological classification, in all the clusters. We therefore have to resort to a statistical treatment of the background contamination. 

While in \cite{Cerulo_2016} we followed an approach based on directly subtracting the field contribution from the observed number counts, for this work we adopted a more robust strategy based on statistical inference, which simultaneously takes into account the contribution to the morphological fractions originating from the cluster (signal) and the background (noise). The full mathematical derivation is presented in \cite{Dagostini_2004}, and we refer to that paper for the details (see also \citealt{Cameron_2011}). Astrophysical applications of this general method can be found in \cite{Andreon_2006_Butcher_Oemler} and \cite{Li_2012}. This method is shown to be more robust with respect to the direct subtraction of field galaxies as it circumvents problems that arise when dealing with small numbers (e.g.\ background-subtracted number counts less than 0).

We built a sample of morphologically classified field galaxies using the ACS F850LP images of the Great Observatories Origins Deep Survey (GOODS, \citealt{Giavalisco_2004}). For each cluster we selected random GOODS galaxies spanning the same colour range of the $(i_{775} - z_{850})$ vs $z_{850}$ cluster red sequence, and with $z_{850} < 24.0$ mag. We classified these galaxies adopting the same procedure described in Section 2.4 and produced 9 control samples. Due to the variable magnitude completeness limits and colour ranges of red sequence galaxies, the latter resulting from the colour cut discussed in Section 2.1, the control morphological samples included numbers of galaxies that varied between a minimum of 156 (XMM1229 control sample) and 424 (RX0152 control sample)\footnote{In \cite{Cerulo_2016} the average cosmic variance in the GOODS fields was shown to amount to 6\%.}.

We estimated the fraction $F_T$ of cluster galaxies of morphological type $T$ as a function of absolute magnitude and stellar mass by building a probability distribution (posterior) that takes into account the contamination from field galaxies. In particular, the best estimate of the fraction was defined as the median of the posterior probability distribution, while the uncertainties were defined as the 16\% and 84\% credible intervals of the distribution. The control fields were used to infer the contamination of the morphological fractions due to field background and foreground galaxies.

In the two low-redshift samples, WINGS and EDisCS, we only considered spectroscopically confirmed cluster members and, for this reason, we treated the measurement of the morphological fractions without considering background contamination. In both cases we did take the spectroscopic incompleteness of the two samples into account and estimated the errors as in \cite{Cameron_2011}.

Following \cite{Cerulo_2014}, we used V-band AB absolute magnitudes ($M_V$) to study morphology as a function of luminosity. We converted the photometry in the observer frame to absolute magnitudes following the method discussed in Appendix B of \cite{Mei_2009} and explained in \cite{Cerulo_2016}. We adopted the same procedure to convert the observer-frame photometry (V and I band magnitudes) in EDisCS to the rest-frame $V_{AB}$ band. Rest-frame $V_{AB}$ magnitudes for the WINGS galaxies were obtained by using the distance moduli provided in the WINGS catalogues and the k-corrections of \cite{Poggianti_1997}. V-band absolute magnitudes in HCS, WINGS, and EDisCS were passively evolved to $z=0$ to enable the comparison between the two samples. The red sequence morphological fractions as a function of absolute magnitude $M_V$ and stellar mass $M_*$ are shown in Figure \ref{fig5} and will be discussed in Section 4{\footnote{The conversions to absolute magnitudes in HCS and EDisCS and the passive evolution corrections for all the samples were performed with the EzGAL Python package, see: \url{http://www.baryons.org/ezgal/}}}.

\subsection{The Morphological Composition of the Red Sequence}

\begin{figure*}
        \subfloat[]{\label{fig:a}}\includegraphics[width=0.4\textwidth, trim=0.0cm 1.5cm 0.0cm 0.0cm, clip, page=1]{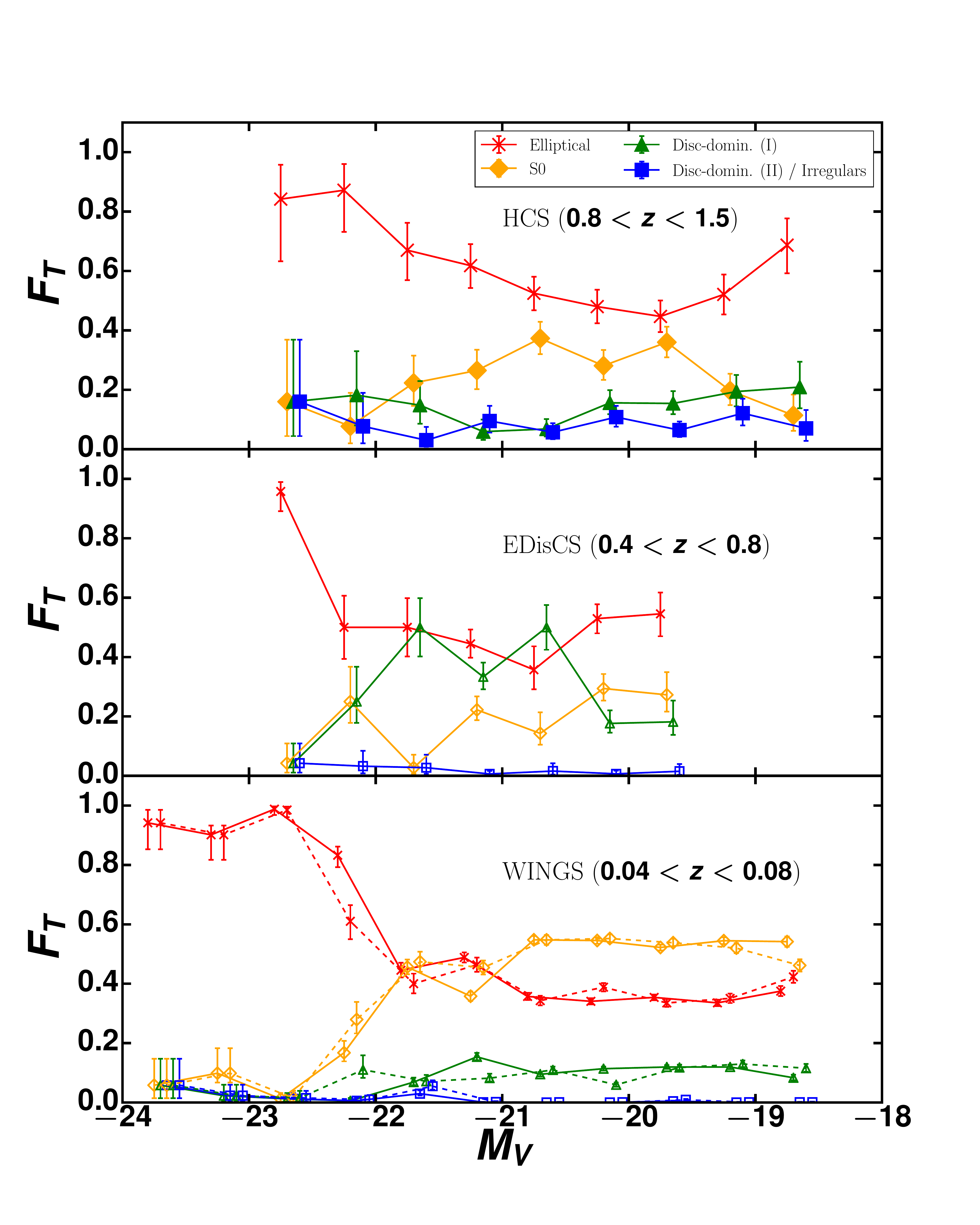}
        \hspace{1mm}
	\subfloat[]{\label{fig:b}}\includegraphics[width=0.4\textwidth, trim=0.0cm 1.5cm 0.0cm 0.0cm, clip, page=2]{plot_RS_morphology_fractions_HCS_Dag2004}
	\caption{{\itshape{Left}}: Background corrected morphological fractions as a function of $V_{AB}$ absolute magnitude along the cluster red sequence in HCS (top panel), EDisCS (middle panel) and WINGS (bottom panel). {\itshape{Right}}: Background corrected morphological fractions as a function of stellar mass along the cluster red sequence in HCS (top panel) and WINGS (bottom panel). The vertical dotted lines at $\log(M_*/M_\odot) = 10.95$ and $\log(M_*/M_\odot) = 11.5$ represent the stellar mass limit of the HCS morphological sample and the maximum stellar mass of HCS red sequence galaxies, respectively. The plots show that elliptical galaxies are the dominant morphological class in the HCS clusters at all luminosities and masses while the red sequence of the WINGS clusters is dominated by ellipticals at $M_V < -21.0$ mag ($\log(M_*/M_\odot) > 11.3$) and by S0s at $M_V > -21.0$ mag ($\log(M_*/M_\odot) < 11.3$). Disc-dominated galaxies make up to 40\% of the red sequence population in the 4 EDisCS clusters considered in this paper. We did not derive stellar masses for EDisCS as the publicly available photometry is not sufficient to determine stellar masses through SED fitting (see Section 2.3). Dotted lines connecting symbols in the bottom-left panel represent the WINGS morphological fractions estimated within a $0.27 \times R_{200}$ radius from the centre of each WINGS cluster.}
\label{fig5}
\end{figure*}

The left-hand panel of Figure \ref{fig5} shows the morphological fractions of red sequence galaxies as a function of $V_{AB}$ absolute magnitude in the HCS (top), EDisCS (middle), and WINGS (bottom) composite red sequences. The right-hand panels of Figure \ref{fig5} show the plots of the morphological fractions as a function of galaxy stellar mass in HCS (top) and WINGS (bottom). It can be seen that the HCS stellar mass completeness limit $\log(M_{lim}/M_\odot) = 10.9$ results in a shallow sample and allows the study of the trends of morphological fractions only within a narrow 0.6 dex mass range. In the following sections we will consider the trends in both the mass selected and the magnitude selected samples with the caveat that the results at masses $\log(M_*/M_\odot) < 10.9$ in the latter sample are valid only at $z<1.46$.

The comparison between HCS and WINGS shows, in agreement with \cite{Mei_2009}, that the cluster red sequence was already dominated by early-type galaxies at $z \sim 1$. However, while the HCS red sequence is dominated by elliptical galaxies at all stellar masses and luminosities, it can be seen that the fraction of S0 galaxies $F_{S0}$ in WINGS becomes higher than the fraction of elliptical galaxies $F_E$ at magnitudes $M_V > -21.0$ mag, and stellar masses $\log(M_*/M_\odot) < 11.3$. We also note an upturn of $F_E$ in HCS at magnitudes $M_V > -19.5$ mag. 

In Section 2.4.3 we pointed out that the contamination from misclassified face-on S0 galaxies should amount to $< 10$\%, suggesting that the upturn is real and not a selection effect. However, we also stressed that faint S0 galaxies, visually and photometrically indistinguishable from ellipticals of the same luminosities, may contaminate the sample of elliptical galaxies in the HCS at $23.0 < z_{850} < 24.0$. In order to detect such a contamination, it is necessary to perform a bulge-disc decomposition, which will be the subject of a forthcoming work. The upturn is also visible, although less pronounced, at stellar masses $\log(M_*/M_\odot) < 10.5$ in the magnitude limited morphological sample. The fractions of early-type disc-dominated, and late-type disc-dominated+Irr galaxies, $F_{D}$ and $F_{late}$, respectively, are low across the entire absolute magnitude and stellar mass ranges spanned by the HCS and WINGS. As shown in Table \ref{table3} late-type galaxies constitute approximately 10-13\% of the red sequence population in both samples.

From Table \ref{table3} we also find that, in agreement with other works (e.g.\ \citealt{Fasano_2001}), there is a $\sim$20\% increase in the total fraction of S0 galaxies towards low redshifts. We find that the difference in $F_{S0}$ between the HCS and WINGS samples is significant at a $6\sigma$ level, suggesting that the cluster red sequence becomes richer in S0 galaxies, which dominate the luminosity distribution at $M_V > M_V^*$ (Figure \ref{fig5} bottom-left panel){\footnote{$M_V^*$ was estimated in \protect{\cite{Cerulo_2016}} for both the HCS and WINGS-SPE red sequence samples. The results, reported in Table 7 of that work, are $M_V^* = -21.08_{-0.02}^{0.08}$ for HCS and $M_V^* = -20.88_{-0.02}^{0.08}$ for WINGS-SPE.}}. Interestingly, the total fraction of early-type galaxies remains approximately constant over redshift. This is in agreement with the results of \cite{Holden_2007} and \cite{van_der_wel_2007}, who studied a sample of clusters in the same halo mass range of HCS, but not with \cite{Vulcani_2011} in the entire EDisCS morphological sample. We note, however, that these two works consider the entire cluster population and do not restrict themselves to the red sequence. The approximately constant fraction of early-type galaxies with redshift with the incresing fraction of S0 galaxies imply that the number of S0 galaxies grows faster than the number of elliptical galaxies. We stress that our data are currently not sufficient to test evolutionary scenarios for galaxies with different morphological types. This topic goes beyond the scope of this paper, which is aimed at presenting the observational differences between the morphological compositions of the red sequence in clusters at $z \sim 1$ and at $z \sim 0$, and will be kept for future work.

\section[Morphological Transformations in Galaxy Clusters]{Discussion}

\subsection{Morphological Transformations in Galaxy Clusters}

The fraction of bulge-dominated galaxies in HCS decreases at faint magnitudes after reaching a maximum at $M_V \sim -20.0$ mag. A similar trend can be observed at $\log(M_*/M_\odot) < 10.8$ in the magnitude limited sample. This result, and the predominance of S0 galaxies in the WINGS red sequence at $M_V > M_V^*$, suggest that the evolutionary paths followed by elliptical and S0 galaxies are different.

The analysis of the morphological fractions that we conducted on the EDisCS sample shows that the red sequence is richer in early-type disc-dominated galaxies with respect to both HCS and WINGS. These galaxies, in the range $-22.0 < M_V < -20.0$, can make up to 40\% of the red population,. This result supports the notion that the excess of S0 galaxies found in WINGS with respect to HCS may be the result of the morphological transformation of quenched spiral galaxies. In \cite{Cerulo_2014} we found a similar result in the MORPHS sample (\citealt{Smail_1997}, \citealt{Dressler_1999}), while \cite{Sanchez_Blazquez_2009} showed that the fraction of late-type galaxies on the red sequence of the full EDisCS sample reached a peak at $0.6 < z < 0.8$. The EDisCS morphological catalogue of \cite{Desai_2007} also contains a {\ttfamily{Dust}} flag which is equal to 1 if the classifiers identified the presence of dust lanes or patches in the HST images. Only 3 of the 22 early-type disc-dominated galaxies in EDisCS have {\ttfamily{Dust}}=1, meaning that only 14\% of the red spirals sample in EDisCS may be red due to the presence of dust.

However, we stress here that the subsample of EDisCS clusters used in the present work is composed of only 4 clusters, and cosmic variance may play an important role. For example, although the Virgo and Hercules clusters have similar velocity dispersions, the Hercules cluster is more spiral-rich than the Virgo cluster. The EDisCS clusters that we use in this paper do not have a uniform spiral fraction on the red sequence. In particular, the clusters cl1216.8-1201 and cl1232.51250 have 13\% and 32\%, respectively, of red spiral galaxies spectroscopically confirmed, while the clusters cl1054.4-1146 and cl1054.7-1245 have 30\% and 43\%, respectively. This points towards the need of having a richer and more uniform morphological sample at the redshifts of the EDisCS clusters.  

We note that the EDisCS spectroscopic sample is 1 mag shallower than HCS and WINGS, and we also note that, as observed in HCS, the two faintest magnitude bins are dominated by elliptical galaxies. As already mentioned in Section 2.4.3, the morphological classification in EDisCS was extensively tested using the light profiles of galaxies; thus the upturn observed in this sample should not be a result of elliptical/S0 misclassification.

The implications of this result are interesting for the study of the different evolutionary paths followed by galaxies with different morphological types; however, we stress that it comes from an analysis conducted on small samples at redshifts $z>0.4$. Although we took into account its incompleteness, the spectroscopic subsample of EDisCS only contains 81 galaxies. Thus we refrain from claiming that the enrichment in spiral galaxies of the red sequence at $0.4 < z < 0.8$ is a general property of all galaxy clusters, and that it constitutes an evidence of the evolutionary link between spiral and S0 galaxies. Instead we only state that our measurements and those presented in \cite{Cerulo_2014} suggest that there may have been morphological transformation of quiescent galaxies from spiral to S0 at redshifts $z<0.4$. In order to test such a scenario, it is necessary to construct a sample of morphologically calssified galaxies at the same redshifts of EDisCS in clusters that are likely descendants of the HCS clusters. Such a task can be currently accomplished with the sample of the Cluster Lensing And Supernova survey with Hubble (CLASH, \citealt{Postman_2012}, \citealt{Rosati_2014}), and we are starting a programme to select galaxies in those clusters and classify them.

From Figure 8 of \cite{Bernardi_2010} one can see that the red sequence in SDSS galaxies at $z<0.1$ is as rich in Sa and Sb galaxies as it is in S0 galaxies. This result is at odd with what we observe in WINGS at similar redshifts. It should be stressed that, while \cite{Bernardi_2010} did not separate into different environments, WINGS is a survey of only galaxy clusters. As shown in Table 5 of \cite{Cerulo_2014}, when we converted the \cite{Fasano_2012} classification into the morphological scheme adopted for HCS we did not include morphologically classified Sa and Sb galaxies in the S0/bulge-dominated class. Thus the subsample of S0 galaxies in WINGS should not be contaminated by Sa and Sb galaxies. One possible explanation of the abundance of red Sa and Sb galaxies observed in  \cite{Bernardi_2010} is therefore that those galaxies may be located, in their majority, in low-density environments.

Although the results presented in this paper do not allow direct evidence of morphological transforation between spiral and S0 galaxies at magnitudes $M_V > -22.0$, this evolutionary hypothesis is supported by two observations, namely the quenching of star formation being more efficient at higher stellar masses (\citealt{Tinsley_1968}, \citealt{Cowie_1996}) regardless of local environment (\citealt{Peng_2010}, \citealt{Rettura_2010, Rettura_2011}, \citealt{Muzzin_2012}), and the evidence for younger stellar populations (0.5-1.0 Gyr) in red sequence S0 galaxies, with respect to ellipticals, in clusters at redshifts $0.8 < z < 1.5$ (\citealt{Tran_2007}, \citealt{Mei_2009}). The first observation implies that the low-luminosity and low-mass S0 galaxies observed in WINGS must have joined the red sequence at redshifts $z < 0.8$, as more time was necessary for them to quench star formation. The second observation not only implies that elliptical and S0 galaxies are different in their stellar populations, but also that S0 galaxies may have a more recent formation history (see also \citealt{Poggianti_2006}). Evidence for an evolutionary link with spiral galaxies in the formation of cluster S0s comes also from the studies of \cite{Sanchez_Blazquez_2009} and \cite{Vulcani_2011} in the EDisCS survey at $0.4 < z < 0.8$. All these authors agree that the total fraction of late-type galaxies (red+blue) increases with redshift, as previously shown in other studies (e.g.\ \citealt{Dressler_1997} \citealt{Couch_1998}, and \citealt{Postman_2005}).

We note that even if we restrict ourselves to the projected area within $0.27 R_{200}$ in WINGS (dashed lines in Figure \ref{fig5}), the trends of the morphological fractions with absolute magnitude do not significantly change. This observation supports the notion that the intermediate-to-low-luminosity S0 galaxies could be already present in the central regions of the clusters at $z \sim 1$ as star-forming spirals which quenched their star formation and then landed on the red sequence at the epochs of the EDisCS clusters. However, the time for a galaxy to cross a distance of 1 Mpc is approximately 1 Gyr \citep{Binney_Tremaine_1987} and, therefore, both the excess of red spirals in EDisCS and the excess of S0 galaxies in WINGS may also be the result of the infalling of star-forming galaxies from the outskirts of the clusters. 

Galaxies that fall towards the cluster centre may have their star formation quenched through ram-pressure stripping or strangulation, or through harassment and tidal interactions with neighbouring galaxies (see \citealt{Treu_2003}). Ram-pressure stripping was shown to be the main responsible for star-formation quenching in spiral galaxies in the Virgo cluster \citep{Gavazzi_2013}. Strangulation, that is the gradual removal of gas from the galaxy halo, has been shown to have time-scales ($\sim 3.0$ Gyr) consistent with the build-up of the red sequence in simulations of galaxy clusters (\citealt{Taranu_2014}, \citealt{Bahe_2015}). \cite{Couch_1998} showed that the majority of S0 galaxies in 3 clusters at $z \sim 0.3$ did not present morphological disturbances or asymmetries. Such a result supports an evolutionary scenario in which S0 galaxies result from the transformation of spiral galaxies quenched through strangulation. \cite{Couch_1998} claimed that only a minority of S0 galaxies could be originated through more violent processes such as galaxy harassment.

\cite{Bekki_2011} showed that tidal interactions with neighbouring galaxies may favour the exhaustion of gas supplies in spiral galaxies by causing the inflow of gas towards their centre and the consequent triggering of starbursts that lead to the growth of their bulges. The end product of this process is an S0 galaxy with a large bulge. \cite{De_Propris_2015} and \cite{De_Propris_2016} showed that red sequence galaxies in clusters at $z>0.8$ tend to have lower S\'{e}rsic indices compared to low-redshift ones, supporting the notion of the growth of bulges in clusters. However, in both works the authors did not divide galaxies into morphological classes thus not allowing one to understand whether the bulge growth involves all the morphological classes or only spiral galaxies. In order to test this scenario we should compare the S\'{e}rsic indices of the red spirals observed in EDisCS and of the S0s observed in WINGS-SPE. Bulge growth triggered by the Bekki-Couch mechanism should happen before the transformation of spiral galaxies into S0s. If the EDisCS red spirals represent one of the phases of the morphological transformation through this process, they should already have a large bulge-to-total ratio. All this information, i.e.\ the S\'{e}rsic indices and bulge-to-total ratio, are currently unavailable in the EDisCS public catalogues, and we leave this analysis for a future work.

One more caveat regarding the Bekki-Couch mechanism is that the morphological transformation was simulated in groups of galaxies, thus haloes with masses ranging from $5 \times 10^{12}$ M$_\odot$ to $1 \times 10^{14}$ M$_\odot$. However, we are considering galaxies that are in the iner region of the clusters (within 700 kpc from the centre). If the EDIsCS red spirals were formed as the result of the tidal interactions of this mechanism, their transformation should have happened in infalling groups (pre-processing). In order to test this scenario, one should investigate the morphological properties of galaxies in the outskirts of galaxies at redshifts higher than EDisCS.

\begin{figure}
       \centering
	\includegraphics[width=0.5\textwidth, trim=2.0cm 30.0cm 0.0cm 6.0cm, clip, page=1]{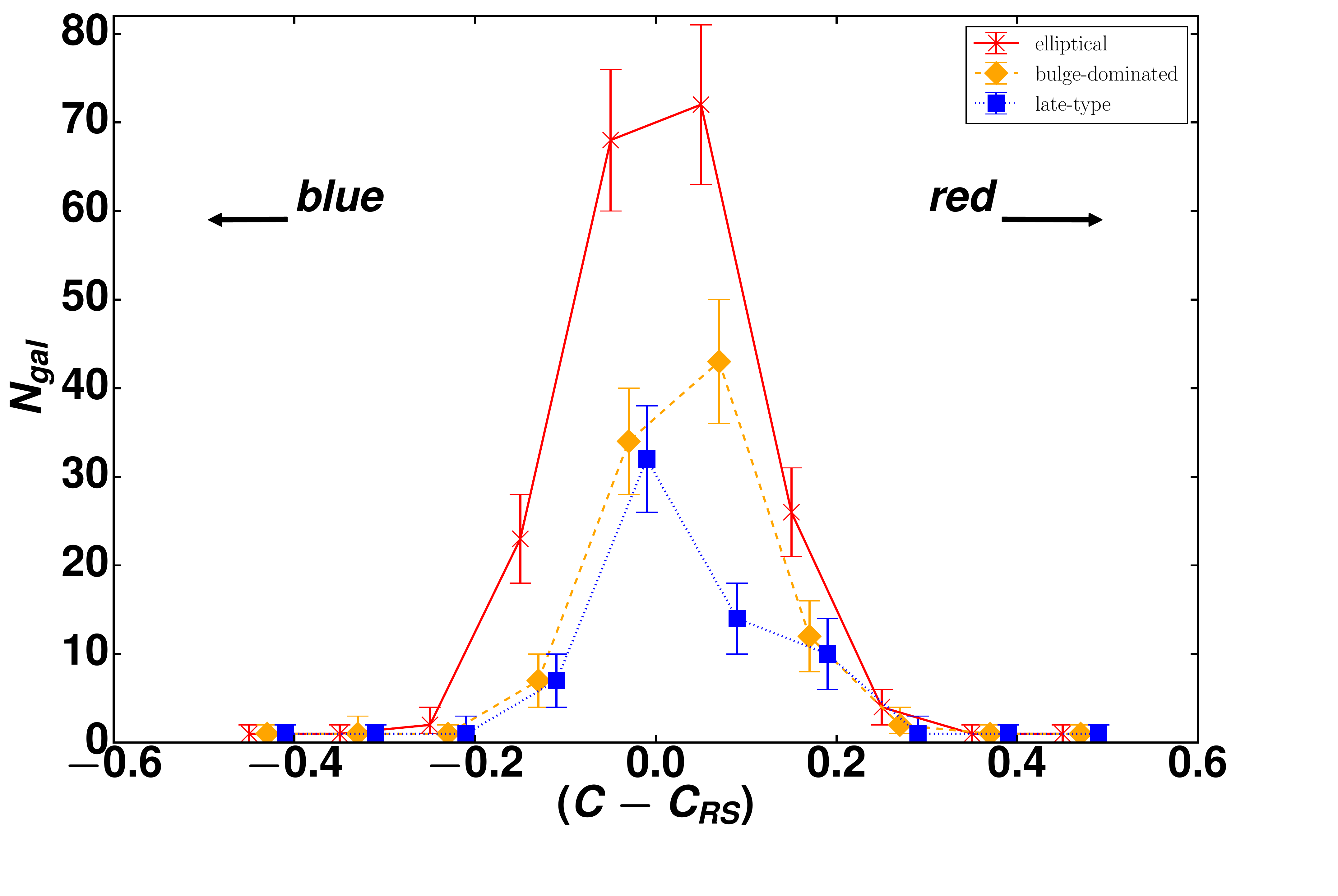}
	\caption{Distributions of galaxy colour measured with respect to the red sequence, $(C-C_{RS})$, for galaxies of different morphological types in HCS. Early-type disc-dominated, late-type disc-dominated and irregular galaxies are grouped into one class of late-type galaxies (blue symbols). Number counts are corrected for field contamination following \protect\cite{Astone_1999}. The arrows in the plot show the directions of colours bluer and redder than the red sequence straight line. We find that late-type galaxies are slightly bluer than early-type galaxies on the red sequence. Although this may suggest that late-type galaxies are younger, we note that the mean colours of early- and late-type galaxies on the red sequence are all consistent (see Section 5.1 for details).}
\label{fig6}
\end{figure}

An indication of the evolutionary link between spiral and S0 galaxies comes also from Table \ref{table4} and Figure \ref{fig6}. The median\footnote{The median values of $(C-C_{RS})$ correspond to the weighted medians of the distributions. This estimator allows one to take into account the effects of field contamination in the HCS sample and the spectroscopic incompleteness in WINGS-SPE.} colours measured with respect to the red sequence for late-type galaxies in HCS and WINGS are $(C-C_{RS}) = -0.01^{+0.16}_{-0.09}$ and $(C-C_{RS}) = -0.03^{+0.06}_{-0.08}$, respectively. These estimates show that late-type galaxies tend to lie on the blue side of the red sequence, although $(C-C_{RS})$ measured for elliptical and S0 galaxies is still consistent with $(C-C_{RS})$ measured for late-type galaxies. This result would suggest that late-type galaxies may be on average younger than ellipticals and S0s, in agreement with the conclusions of \cite{Tran_2007}. However, evidence for differences in the colours of galaxies with different morphological types is very weak, and we are not able to statistically distinguish amongst the colour distributions of red sequence galaxies with different morphologies.

We think that the mechanisms that lead to the consumption of gas and to the consequent fading of spiral arms in disc galaxies, like ram-pressure stripping and strangulation, or to the disruption of spiral arms and/or bulge growth, like harassment or the Bekki-Couch mechanism, are all competing in the dense environment of galaxy clusters and are the main triggers for the morphological transformation of spiral galaxies into S0 galaxies. However, we point out that S0 galaxies may also be originated through mergers (see e.g.: \citealt{Pota_2013} and \citealt{Donofrio_2015}). Interestingly, \cite{Carollo_2016} show that environmental quenching processes do not globally alter the morphology of galaxies in low-redshift groups.

In order to test the role of each of these mechanisms, a more detailed morphological scheme based on the detection of structural disturbances and asymmetries needs to be adopted (see e.g.: \citealt{Neichel_2008}, \citealt{Delgado_Serrano_2010}, \citealt{Nantais_2013a}). Deeper images allowing the study of the morphological composition of the red sequence down to the faint end, at magnitudes $z_{850} > 24.0$, and spectra enabling the study of stellar populations in red sequence galaxies with different morphologies, are also needed in order to test the evolutionary link between S0 and spiral galaxies that we infer from our results. This work is currently ongoing.

\begin{table*}
  \caption{Median ellipticities and total fractions of red sequence galaxies $F_T$ in HCS and WINGS. The uncertainties on the total morphological fractions correspond to the boundaries of the binomial 68\% confidence intervals estimated as in \protect\cite{Dagostini_2004} and \protect\cite{Cameron_2011}. The median values of the ellipticity of WINGS take into account the incompleteness of the spectroscopic sample. Ellipticities are only quoted for early-type disc-dominated galaxies in the two samples (see Section 2.4.3).}
  \begin{minipage}[c]{0.8\textwidth}
    \centering
  \begin{tabular}{|c|c|c|c|c|c|c|}
     \hline
     \multicolumn{1}{|c}{Morphological}  & \multicolumn{1}{c}{Ellipticity} & \multicolumn{1}{c}{$F_T$} & \multicolumn{1}{c}{Ellipticity} & \multicolumn{1}{c|}{$F_T$} \\
     \multicolumn{1}{|c}{Type}  & \multicolumn{1}{c}{(HCS)} & \multicolumn{1}{c}{(HCS)} & \multicolumn{1}{c}{(WINGS)}  & \multicolumn{1}{c}{(WINGS)} \\
     \hline
      \hline
                            &                         &                         &                          &                           \\
      Elliptical            & $0.18^{+0.09}_{-0.11}$    & $0.53_{-0.02}^{+0.02}$    & $0.17^{+0.15}_{-0.12}$     & $0.375_{-0.005}^{+0.005}$    \\
                            &                         &                         &                          &                            \\
      Bulge-Dominated       & $0.38^{+0.10}_{-0.12}$    & $0.27_{-0.02}^{+0.02}$    & $0.47^{+0.16}_{-0.20}$     & $0.513_{-0.005}^{+0.005}$     \\
                            &                         &                         &                          &                             \\
                            &                         &                         &                          &                             \\
      Early-type            & $0.45^{+0.15}_{-0.16}$    & $0.128_{-0.015}^{+0.017}$ & $0.6^{+0.2}_{-0.3}$        & $0.110_{-0.003}^{+0.003}$      \\
      Disc-Dominated        &                         &                         &                          &                              \\
                            &                         &                         &                          &                              \\
      Late-type             &                         &                         & $0.073_{-0.012}^{+0.013}$  &  $0.0023_{-0.0004}^{+0.0006}$   \\
      Disc-Dominated + Irr  &                         &                         &                          &                              \\
                            &                         &                         &                          &                               \\
  \hline
  \end{tabular}
  \end{minipage}
  \label{table3}
\end{table*}

\begin{table*}
  \caption{Median colours relative to the red sequence $(C-C_{RS})$ in HCS and WINGS. The median values of $(C-C_{RS})$ in WINGS take into account the incompleteness of the spectroscopic sample. The median $(C-C_{RS})$ of HCS is corrected for background contamination following \protect\cite{Astone_1999}. In this table, early-type disc-dominated, late-type disc-dominated, and irregular galaxies are grouped together into one class of late-type galaxies.}
  \begin{minipage}[c]{0.8\textwidth}
    \centering
  \begin{tabular}{|c|c|c|}
     \hline
     \multicolumn{1}{|c}{Morphological}  & \multicolumn{1}{c}{$(C-C_{RS})$}  & \multicolumn{1}{c}{$(C-C_{RS})$} \\
     \multicolumn{1}{|c}{Type}  & \multicolumn{1}{c}{(HCS)} & \multicolumn{1}{c}{(WINGS)} \\
     \hline
      \hline
                            &                        &                        \\
      Elliptical            & $0.01^{+0.11}_{- 0.09}$  & $0.00^{+0.07}_{-0.06}$   \\
                            &                        &                        \\
      Bulge-Dominated       & $0.02^{+0.08}_{-0.09}$   & $-0.02^{+0.07}_{-0.06}$  \\
                            &                        &                        \\
      Late-type             & $-0.01^{+0.16}_{-0.09}$  & $-0.03^{+0.06}_{-0.08}$  \\
                            &                         &                        \\
  \hline
  \end{tabular}
  \end{minipage}
  \label{table4}
\end{table*}

\subsection{Morphological Evolution at the Bright End of the Red Sequence}

The bright end of the red sequence in HCS, EDisCS, and WINGS, where most of the brightest cluster galaxies (BCG) reside in all the samples, is dominated by elliptical galaxies which, in the low-z sample, can be as massive as $10^{12.2} M_\odot$. \textcolor{black}{The HCS is not populated by galaxies with masses $\log(M_*/M_\odot) > 11.5$. This result points towards a stellar mass growth higher than the factors estimated for BCGs at $0<z<1$ ( 2-2.5 \citealt{Lidman_2012, Ascaso_2014}), suggesting that the bright red sequence galaxies in WINGS are exceptionally massive.} 

However, we point out that the range $\log(M_*/M_\odot) > 11.5$ is poorly populated in WINGS and corresponds to about 1\% of the full WINGS-SPE sample. This suggests that these galaxies are peculiar objects which may host younger stellar populations than those expected in massive elliptical galaxies.

On the other hand, aperture photometry of nearby massive elliptical galaxies is difficult because the light distributions of these objects have extended wings. As discussed in \cite{von_der_linden_2007}, a significant fraction of light can be lost in these galaxies even with large apertures, leading to the underastimation of their flux (see also \citealt{Bernardi_2013}). Thus the high stellar masses may also be the results of an incorrect treatment of the most luminous galaxies, whose photometry was performed together with all the other galaxies in the sample.

Interestingly, \cite{Bernardi_2011a, Bernardi_2011b} found that the colour vs magnitude, colour vs size, and colour vs stellar mass relationships of early-type galaxies in the SDSS curve upwards at stellar masses $M_* > 2 \times 10^{11} M_\odot$. The authors suggest that this stellar mass represents a transition value above which the formation history of early-type galaxies and in particular the build-up of their stellar mass is dominated by major dry mergers. In a later work, \cite{Cappellari_2013_ATLAS3D_Paper_XX} showed that a similar stellar mass represented the value above which the population of early-type galaxies in the ATLAS$^{3D}$ survey \citep{Cappellari_2011_ATLAS3D_Paper_I} is dominated by pressure-supported slow rotators. Similar conclusions can also be found in \cite{van_der_Wel_2009} and \cite{Mitsuda_2017}. Such mass corresponds to the $\log(M_*/M_\odot) = 11.3$ above which we find that the WINGS red sequence is dominated by elliptical galaxies which, as shown in \cite{Cappellari_2011} and \cite{Cappellari_2013_ATLAS3D_Paper_XX} are mostly slow-rotators without discs.

\subsection{The Comparison with XMM1229}

The results presented in this paper do not agree with those shown in \cite{Cerulo_2014} for the morphological fractions in XMM1229. In fact, in that cluster we found that, while the fraction of elliptical galaxies remained approximately constant along the red sequence, the fraction of S0 galaxies reached a maximum at $M_V \sim -20.8$ mag and then decreased. Correspondingly, the fraction of disc-dominated galaxies increased at $M_V > -20.3$ mag. This suggested that the faint red disc-dominated galaxies could be the progenitors of the S0 galaxies dominating the red sequence of the WINGS clusters at the same magnitudes.

The trends observed in XMM1229 may be peculiar to that cluster, although we stress that the magnitude limit of the morphological sample $z_{850} = 24.0$ mag produces a variable sampling of each cluster's red sequence. For example, as shown in Figure \ref{fig1}, in the clusters RCS2319 and RDCS1252, the magnitude selection does not allow us to investigate the faint end of the red sequence down to the 90\% completeness limit (diagonal dashed line), while in other clusters, such as XMM1229 or RCS2345, we are able to classify a larger fraction of the cluster red sequence. Therefore, we do not exclude that with deeper images one may detect a similar upturn in the fraction of late-type galaxies at the faint end of the red sequence.

\section{Summary and Conclusions}

We have analysed the morphological composition of red sequence galaxies in a sample of clusters at $0.8 < z < 1.5$ drawn from the HCS data-set. We have investigated background-corrected morphological fractions as a function of galaxy luminosity and stellar mass, and we have compared with lower-redshift samples of clusters selected from the spectroscopic follow-up observatios of the WINGS and EDisCS surveys. We have used the available HST/ACS F850LP images to classify galaxies in the HCS clusters, selecting all red sequence galaxies with $z_{850} < 24.0$ mag. We have subdivided galaxies in HCS, EDisCS, and WINGS in elliptical, S0, early-type disc-dominated, late-type disc-dominated and irregular.

The red sequence in clusters at high and low redshifts is dominated by early-type galaxies, which comprise $\sim$85\% of the total red sequence population. While the HCS red sequence is dominated by elliptical galaxies at all luminosities and stellar masses, the WINGS red sequence becomes dominated by S0 galaxies at $M_V > M_V^*$. At luminosities brighter than this value the red sequence is dominated by elliptical galaxies, which reach stellar masses as high as $10^{12.2} \mbox{ } M_\odot$. The red sequence undergoes a $\sim$20\% enrichment in S0 galaxies towards lower redshfits. We also observe that late-type galaxies make up only 10-13\% of the red sequence population at both $z \sim 1$ and $z \sim 0.05$, but can constitute up to 40\% of the red sequence in clusters at $0.4 < z < 0.8$.

The differences between the morphological compositions of low- and high-redshift red sequences, with the enrichment in S0 galaxies on the low-redshift red sequence, imply that elliptical and S0 galaxies follow different evolutionary paths. The enrichment in disc-dominated galaxies observed on the red sequence of the 4 EDisCS clusters studied in this paper suggests that, at least in the range $-22.0 < M_V < -20.0$, the excess of S0 galaxies observed at the same magnitudes in WINGS may be the result of morphological transformation of passive spiral galaxies. 

We stress that our conclusions are based on small samples of clusters at $z>0.4$ and that larger samples are required in order to constrain the morphologcal properties of galaxies at these redshifts. We also point out that deeper images and spectra are needed in order to study the morphology of galaxies down to the faint end of the red sequence and to complement the study of galaxy morphology with that of stellar populations. We finally remind that the magnitude selection limit $z_{850} = 24.0$ mag results in a non-uniform sampling of the red sequences of individual clusters and that in some clusters (e.g.\ RDCS1252) we are not able to classify galaxies at the faint end of the red sequence.

Despite all these limitations, the present work provides tantalising evidence that makes us better visualise the pathway for the formation of S0 galaxies in high-density environments along cosmic time.

\section*{Acknowledgements}

We thank the anonymous referee for the helpful and constructive comments. The results presented in this paper expand one chapter of P.C.’s PhD thesis. We thank Michael Brown, Casey Papovich, and Ray Sharples for their constructive feedback and suggestions. This work was performed on the gSTAR national facility at Swinburne University of Technology. gSTAR is funded by Swinburne and the Australian Governments Education Investment Fund.We thank the Swinburne supercomputing and ITS teams for their support. We would like to thank Bianca Maria Poggianti and Alessia Moretti for providing us with the latest versions of the WINGS catalogues. We also thank Boris H\"{a}u{\ss}ler for providing the latest up to date scripts for galaxy image simulations. We thank Julie B. Nantais for providing updated redshift catalogues for the cluster RDCS J1252.9-2927. We thank Yara Jaff\'{e} for providing suggestions and stimulating discussions on the use of the EDisCS and WINGS data. P.C. acknowledges the support provided by FONDECYT postdoctoral research grant no 3160375, by a Swinburne Chancellor Research Scholarship and an AAO PhD Scholarship. W.J.C. gratefully acknowledges the financial support of an Australian Research Council Discovery Project grant throughout the course of this work. R.D. gratefully acknowledges the support provided by the BASAL Center for Astrophysics and Associated Technologies (CATA), and by FONDECYT grant No. 1130528. The data in this paper were based in part on observations obtained at the ESO Paranal Observatory (ESO programme 084.A-0214). Based on observations obtained at the Gemini Observatory, which is operated by the Association of Universities for Research in Astronomy, Inc., under a cooperative agreement with the NSF on behalf of the Gemini partnership: the National Science Foundation (United States), the National Research Council (Canada), CONICYT (Chile), Ministerio de Ciencia, Tecnolog\'{i}a e Innovaci\'{o}n Productiva (Argentina), and Minist\'{e}rio da Ci\^{e}ncia, Tecnologia e Inova\c{c}\~{a}o (Brazil). Based on observations made with the NASA/ESA Hubble Space Telescope, obtained from the data archive at the Space Telescope Science Institute. STScI is operated by the Association of Universities for Research in Astronomy, Inc. under NASA contract NAS 5-26555. These observations are associated with Program ID 12051. Based on data obtained from the ESO Science Archive Facility (Program ID 073.A-0737(A)). Part of the data presented herein were obtained at the W.M. Keck Observatory, which is operated as a scientific partnership among the California Institute of Technology, the University of California and the National Aeronautics and Space Administration. The Observatory was made possible by the generous financial support of the W.M. Keck Foundation. The authors wish to recognize and acknowledge the very significant cultural role and reverence that the summit of Mauna Kea has always had within the indigenous Hawaiian community.  We are most fortunate to have the opportunity to conduct observations from this mountain.

\bibliographystyle{mn2e}
\small
\itemindent -0.48cm
\bibliography{biblio}

\bsp

\appendix

\section{Morphological Classification of Red Sequence Galaxies in the HAWK-I Cluster Survey (HCS)}

In the following figures we plot the postage stamp images of morphologically classified red sequence galaxies in the cluster RDCS J1252.9-2927 (RDCS1252 $z=1.24$). The images are cutouts of the ACS F850LP images of the cluster field. The size of each image is 4.05$''$ on each side. This angular projected size corresponds to a physical projected size of approximately 34 kpc. In each postage stamp image North is up and East is to the left.

\newpage
\begin{figure*}
  \centering
  \subfloat[]{\label{fig:a}}\includegraphics[width=0.8\textwidth, trim=0.0cm 0.0cm 0.0cm 0.0cm, clip, page=1]{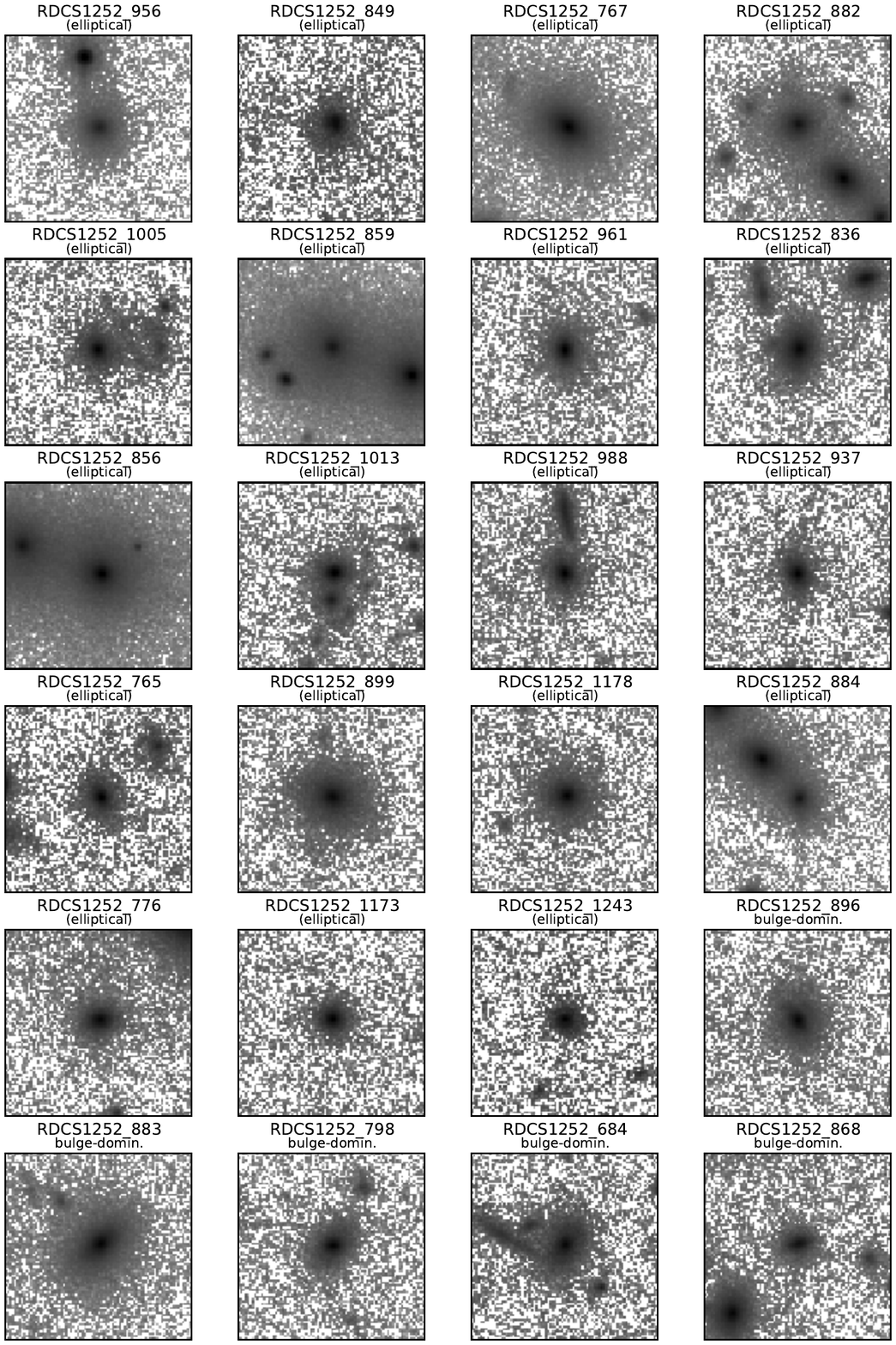}
	\caption{Postage stamp images of the morphological sample of the cluster RDCS J1252.9-2927 (RDCS1252) at $z=1.24$}
\end{figure*}

\begin{figure*}
  \ContinuedFloat 
  \centering 
  \subfloat[]{\label{fig:b}}\includegraphics[width=0.8\textwidth, trim=0.0cm 16.0cm 0.0cm 0.0cm, clip, page=2]{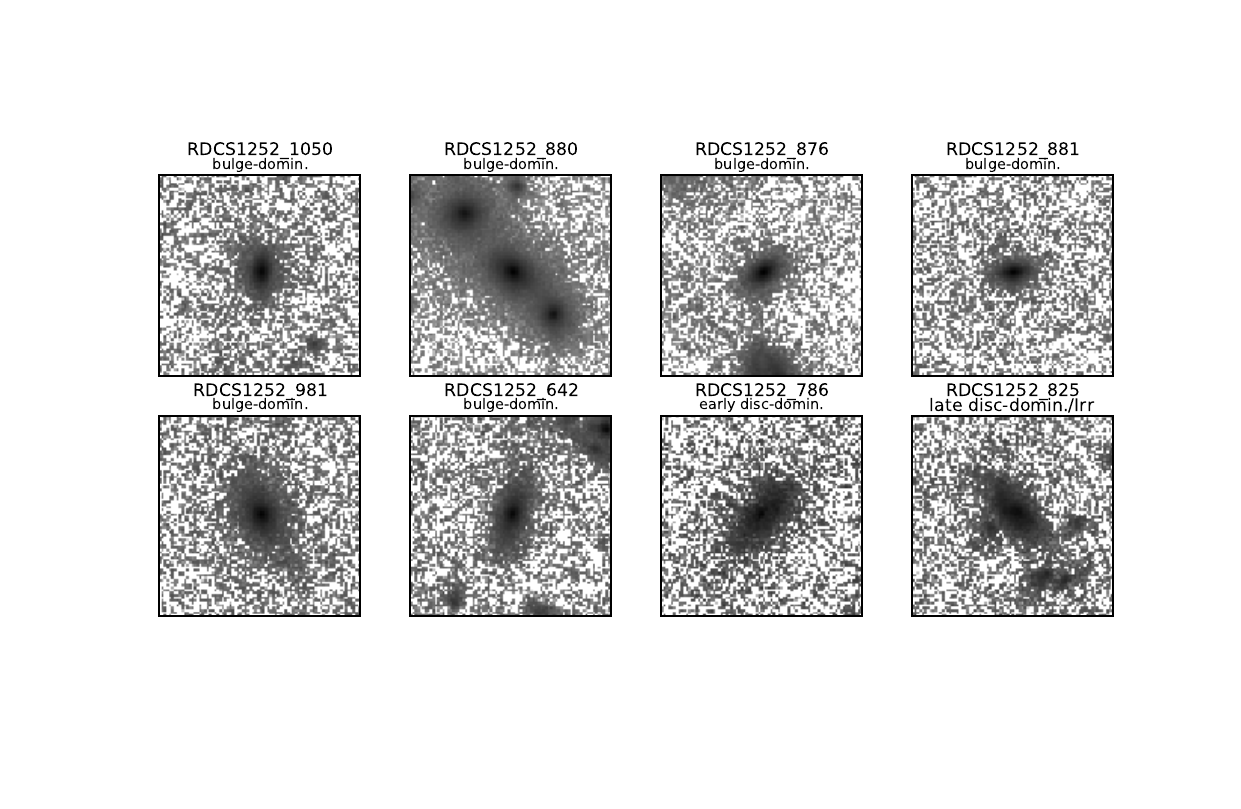}
  \caption{Continued.}
\end{figure*}

\label{lastpage}

\end{document}